\documentclass[useAMS,usenatbib]{mn2e}

\include{epsf}
\usepackage{subfigure}
\usepackage{graphicx}
\usepackage{longtable}
\usepackage{rotating}
\usepackage{amssymb}

\title[Clustering of 2PIGG galaxy groups with 2dFGRS galaxies]{Clustering of 2PIGG galaxy groups with 2dFGRS galaxies}
\author[G. Mountrichas \& T. Shanks]{G. Mountrichas \& T. Shanks}

\begin{document}

\date{9 April 2007}

\pagerange{\pageref{firstpage}--\pageref{lastpage}} \pubyear{2007}

\maketitle

\label{firstpage}

\begin{abstract}

Prompted by indications from QSO lensing that there may be more
mass associated with galaxy groups than expected from virial analyses, we
have made new dynamical infall estimates of the masses associated with
2PIGG groups and clusters. We have analysed the redshift distortions in
the cluster-galaxy cross-correlation function as a function of cluster
membership, cross-correlating $z<0.12$ 2PIGG clusters and groups with the
full 2dF galaxy catalogue. We have made estimates of the dynamical infall parameter $\beta\propto\frac{1}{b}$ where b is the bias parameter and new estimates of the group velocity dispersions for group membership classes
out to $z<0.12$. We first find that, out to 30-40h$^{-1}$Mpc, the
amplitude of the full 3-D redshift space cross-correlation function,
$\xi_{cg}$, rises monotonically with group membership. We use a simple
linear-theory infall model together with a Gaussian velocity dispersion to
fit $\xi(\sigma,\pi)$ in the range $5<s<40$h$^{-1}$Mpc. We find that the $\beta$ versus membership relation for the data shows a minimum
at intermediate group membership $n\approx20$ or
$L\approx2\times10^{11}$h$^{-2}$L$_{\sun}$, implying that the bias and hence $M/L$ ratios rise by a significant factor ($\approx5\times$) both for small
groups and rich clusters. The minimum for the mocks is at a $2-3\times$
lower luminosity than for the data. However, the mocks also show a
systematic shift between the location of the $\beta$ minimum and the M/L minimum at $L\approx10^{10}$h$^{-2}$L$_{\sun}$ given by direct calculation using the known DM
distribution. We
consider physical reasons for this difference. Our overall conclusion is
that bias estimates from dynamical infall appear to support the minimum in
star-formation efficiency at intermediate halo masses. Nevertheless, there may still be
significant systematic problems arising from measuring $\beta\propto\frac{1}{b}=\delta \rho _{mass}/\delta \rho _{galaxies}$ using
large-scale infall rather than M/L using small-scale velocity dispersions.

\end{abstract}


\section{Introduction}

Previous QSO lensing results (Myers et al. 2003, 2005, Mountrichas \& Shanks 2007) have indicated that galaxies are anti$-$biased on small scales at a higher level than predicted by the standard cosmological model. In continuation of this investigation, we now make dynamical infall estimates of the masses associated with 2PIGG groups and clusters to compare our results with QSO lensing estimates and also the velocity dispersion mass estimates of Eke et al. 2006. 

Here we obtain those dynamical infall estimates of group masses using redshift distortion analysis. The analytical background is based on the fact that the observed redshifts of objects at cosmological distances do not reflect their true distances as they are affected by the expansion rate of the universe (Hubble flow) and their peculiar velocities (velocity dispersion) with respect to their local rest frame. These two effects distort the clustering pattern of the objects in two ways: at small scales the large velocity dispersion causes an elongation in the redshift direction, along the line-of-sight (Fingers of god) and on large scales the infall of matter into higher density regions causes a flattening of the clustering pattern across the line-of-sight. Geometrical distortions are also introduced if a wrong cosmological model is assumed. 

These redshift-space distortions can be investigated using either the spatial two-point correlation function (e.g. Ratcliffe et al 1997) or its Fourier transform, the power spectrum (e.g. Outram, Hoyle and Shanks 2000). The objects for which the clustering pattern is studied can be Lyman Break Galaxies (LBGs, da $\hat{A}$ngela et al. 2005), QSOs (da $\hat{A}$ngela et al. 2006), Luminous Red Galaxies (LRGs, Ross et al. 2006), QSO-LRG (Mountrichas \& Shanks 2007, in prep) and other galaxies or galaxy groups. In recent years many galaxy catalogues have been used for this purpose, the Durham/UKST Galaxy Redshift Survey (Ratcliffe et al. 1996), the IRAS Point Source Redshift Survey (PSCz, Taylor et al. 2000), the Nearby Optical Galaxy (NOG) sample (Giuricin et al. 2001), the Zwicky catalogue (Padilla et al. 2001) and others. The parameters we can measure in these analyses are the velocity dispersion of the galaxies (or galaxy groups), the infall parameter $\beta$, the amplitude of galaxy clustering $r_0$, the density parameter, $\Omega_m$, and the bias factor, b.   

Recently, large group catalogues have been used to study the variation of their halo mass-to-light ratio (M/L) with luminosity. These calculations show us in what halo sizes mass is most efficiently converted into stars. Most recently, the 2PIGG team analysed 2PIGG group velocity dispersion on the 2PIGG group catalogue (Eke et al. 2004a) and found a variation of the halo mass-to-light ratio with group luminosity. More precisely, they found an increase by a factor of 5 in the $b_J$ band and 3.5 in the $r_F$ band, of the halo M/L with a 100 times increased group luminosity (Eke et al. 2004b) as well as a minimum at a total group luminosity of $L\approx5\times10^9$h$^{-2}$L$_{\sun}$ (Eke et al. 2006). Padilla et al. 2004 measured the clustering amplitude of 2PIGG galaxy groups. They found that the most luminous groups are 10 times more clustered than the full 2PIGG catalogue. 

In the work presented here, we try to estimate group masses using dynamical infall rather than just the velocity dispersion. We believe that estimating masses by infall may be more accurate because no stability or virial assumptions are needed. We do this by estimating the 3-D redshift space cross-correlation function, $\xi _{cg} (s)$, between the 2PIGG groups and clusters with the 2dFGRS galaxies (Section 3) as well as the semi-projected cross-correlation function, $w_p(\sigma)/ \sigma$ and the real-space cross-correlation function, $\xi _{cg}(r)$ (Sections 4, 5).  In Section 6 we present our $\xi _{cg}(\sigma, \pi)$ results and constrain the cluster-galaxy infall parameter, $\beta$, by modelling the redshift-space distortions. In Section 7 we find the group luminosities and we repeat our previous measurements, as a function of luminosity. Finally, our conclusions are presented in Section 8.   
  
\begin{table}
\caption{Number of groups in the data and mock catalogues ($z<0.12$).}
\centering
\setlength{\tabcolsep}{4.0mm}
\begin{tabular}{lcc}
       \hline
$n_{gal} $ & data & mock \\
       \hline \hline
=1  & $45,227$ & $20,152$ \\
       \hline
2-3 & $11,742$  & $9,959$ \\
       \hline
=4 & $2,517$  & $1,096$  \\
       \hline
5-8 & $3,005$  & $1,375$ \\
       \hline
9-17  & $1,082$  & $592$ \\
       \hline
18-29 & $204$  & $201$  \\
       \hline 
30-44 & $54$  & $100$ \\
       \hline 
45-69 & $43$  & $67$  \\
       \hline 
$\geq70$ & $39$  & $80$  \\
       \hline
\label{table:numbers}
\end{tabular}
\end{table}

\begin{figure*}
\begin{center}
\centerline{\epsfxsize = 9cm
\epsfbox{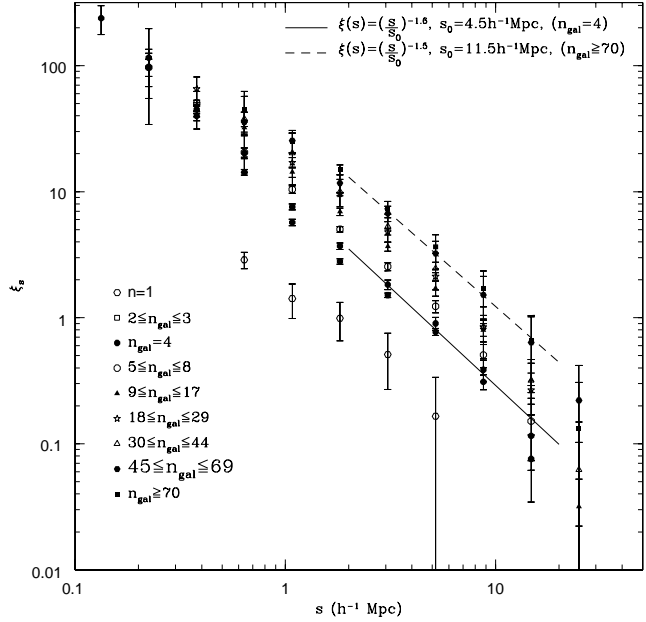}}
\caption{The galaxy group-galaxy redshift-space cross-correlation function $\xi _{cg}(s)$ for different group memberships. The solid line is the best fit to our $\xi _{cg}(s)$ measurements for our group sample with $n_{gal}=4$. The slope is $\gamma=1.6$ and the correlation length is $s_0=4.5$h$^{-1}$Mpc. The dashed line is the fit to our groups with the largest membership, i.e. $n_{gal}\geq 70$, with $\gamma=1.5$ and $s_0=11.5$h$^{-1}$Mpc.}
\label{fig:xis}
\end{center}
\end{figure*}

\begin{figure*}
\begin{center}
\centerline{\epsfxsize = 9cm
\epsfbox{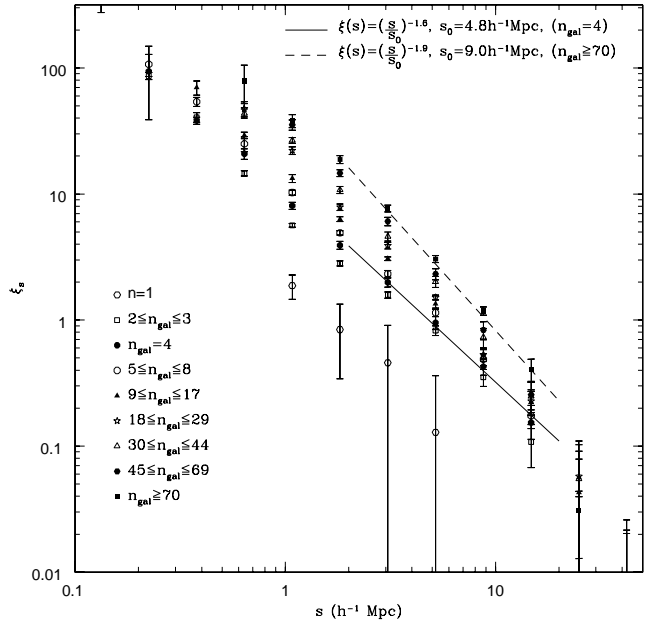}}
\caption{The galaxy group-galaxy redshift-space cross-correlation function $\xi _{cg}(s)$ for different group membership using the mock catalogues. The solid line is the best fit to our $\xi _{cg}(s)$ measurements for our group sample with $n_{gal}=4$. The slope is $\gamma=1.6$ and the correlation length is $s_0=4.8$h$^{-1}$Mpc. The dashed line is the fit to our groups with the largest membership, i.e. $n_{gal}\geq 70$, with $\gamma=1.9$ and $s_0=9.0$h$^{-1}$Mpc.}
\label{fig:xis_mock}
\end{center}
\end{figure*}

\section{Data and Clustering Analysis}

\subsection{Data}

The 2dFGRS galaxies (Lewis et al. 2002) are here used both as the source of the group and cluster catalogue and as dynamical tracers. The 2dFGRS sample contains 191,440 galaxies (NGP$+$SGP) once cuts have been applied for field and sector completeness. The magnitude limit of the survey is $b_J=19.45$. 

The galaxy groups we use are taken from the 2dFGRS Percolation-Inferred Galaxy Group (2PIGG) catalogue (Eke et al. 2004a). The groups are derived using a friends-of-friends (FOF) algorithm. This algorithm has been calibrated and tested using mock 2dFGRS catalogues and then applied to the real 2dFGRS in order to construct the 2PIGG catalogue. The algorithm as well as its calibration is described in detail by Eke et al. 2004a. The catalogue consists of 28,877 groups with at least two members (in the whole redshift range). Following Eke et al. we shall only use groups with $z<0.12$ because at higher redshifts the fraction of the total group luminosity observed falls below 50$\%$ and  also the contamination rate increases. We also follow the group membership classes used in Fig. 5 of Eke et al. 2004a but we split the $18\leq n_{gal}\leq 44$ and $n_{gal}\geq 45$ each into 2 subsamples to get group samples with intermediate and high memberships. Moreover we use two more $`$group' samples at small memberships, those with $2\leq n_{gal}\leq 3$ and $n=1$. The numbers of each group sample used are shown in Table \ref{table:numbers}.

In our measurements we also use mock catalogues constructed by the 2PIGG team (Eke et al. 2004a) and which are available on the World Wide Web. High-resolution N-body simulations of cosmological volumes (Jenkins et al. 1998) and a semi-analytical model of galaxy formation (Cole et al. 2000) were used for the construction of the mock catalogues. A full description of the catalogues and the methods used for their construction is given by Eke et al. 2004a. The mock catalogue consists of 25,201 groups in both the NGP and the SGP (in the whole redshift range). This number is slightly lower than the corresponding one from the real data. We divide the mock groups in the same classes as for the data and apply the redshift cut of $z<0.12$. The numbers are again shown in Table \ref{table:numbers}. We note that the mock catalogue contains more large membership groups than the data catalogue.

For our random catalogues, field and sector completeness have been taken into account. The catalogues consist of $\approx11\times$ the number of our galaxies, i.e. 2,105,840 random points in the NGP$+$SGP.

\subsection{Cross-correlation and errors estimators}

The correlation function estimator that we use for our analysis is  

\begin{equation}
w(\theta )=\frac{DD(\theta )}{DR(\theta )}\frac{N_{rd}}{N_{gal}}-1
\label{eqn:wtheta}
\end{equation}
where $N_{rd}$ is the number of the random points in our catalogue, $N_{gal}$ is the number of galaxies we use, $DD(\theta)$ is the data-data pairs, i.e. group-galaxy pairs and $DR(\theta)$ is the group-random point pairs counted at angular separation $\theta$. 

The errors we use throughout our analysis are Field-to-Field errors. For that purpose we have divided NGC and the SGC each into 3 equal areas and then measure the correlation functions for each one of these 6 areas. The Field-to-Field error is then given by the following expression

\begin{equation}
\sigma_\omega^2(\theta) =
\frac{1}{N-1}\sum_{L=1}^{N}\frac{DR_L(\theta)}{DR(\theta)}[\omega_L(\theta)-\omega(\theta)]^2
\end {equation}
where $N=6$, $DR_L(\theta)$ is the data-random pairs in the subarea, $DR(\theta)$ is the overall number of data-random pairs, $\omega_L(\theta)$ is the correlation function measured in the subarea and $\omega(\theta)$ is the overall correlation function. In the next Section we shall see how the Field-to-Field errors compare with Poisson errors ($\sigma(\theta)=\frac{\sqrt{DD(\theta)}}{DR(\theta)}$) for our $\xi (s)$ measurements.

\begin{figure}
\begin{center}
\centerline{\epsfxsize = 7.0cm
\epsfbox{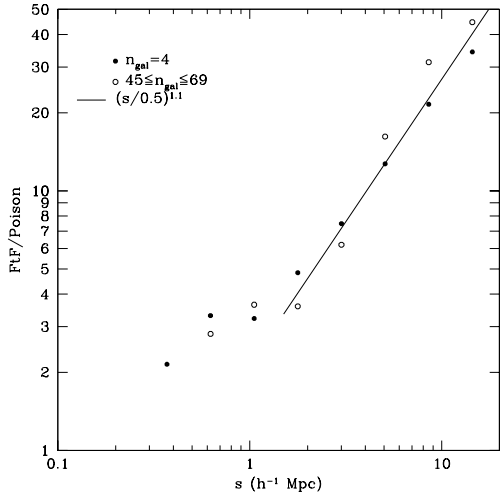}}
\caption{Variation between $\xi _{cg}(s)$ Field-to-Field and Poisson errors over the separation s. Filled circles correspond to galaxy groups with $n_{gal}=4$ and open circles to galaxy groups with $45\leq n_{gal}\leq 69$. On our scales of interest (i.e. $2-20$h$^{-1}$Mpc) the results follow a power law, i.e. $(s/0.5)^{1.1}$.}
\label{fig:field_over_poisson2}
\end{center}
\end{figure}

\section{Redshift$-$space cross-correlation function}

In this Section we present our measurements for the group-galaxy redshift-space cross-correlation function $\xi _{cg}(s)$ for different group memberships, as they were described in Section 2. The purpose of these measurements is to study how the clustering amplitude changes for different groups and clusters and to obtain the values for the correlation amplitude $s_0$ and the slope $\gamma$ from single power law fits, in order to use them later on when we shall model our $\xi (\sigma, \pi)$ measurements (even though, as shall be explained, we will ultimately let $s_0$ float as a free parameter in the redshift distortion fits). We first measure $\xi _{cg}(s)$ in each Galactic Cap separately, i.e. NGC and SGC and then by adding the DDs together and the normalised DRs and using the expression \ref{eqn:wtheta} we combine these measurements to get the overall result. We should note that the SGC tends to give a slightly bigger correlation length, which is in accordance with the fact that larger membership clusters are found in the SGC. The combined results (North$+$South) are shown in Fig. \ref{fig:xis}. 

As expected the groups with larger memberships have higher correlation lengths ($s_0$) and the ungrouped galaxies have by far the smallest correlation length. This is reflected in the fits ($2\leq s\leq 20$h$^{-1}$Mpc) to these cross-correlation functions. All the fits assume the same power law form, i.e. $\xi _{cg}(s)=(s/s_0)^{-\gamma}$ with both $s_0$ and $\gamma$ allowed to vary. The error estimates come from the 1$\sigma$ deviation from the minimized $\chi ^2$ on these fits. The solid line in Figure \ref{fig:xis} is the best fit to our $\xi _{cg}(s)$ measurements for our group sample with $n_{gal}=4$. The slope is $\gamma=1.6\pm0.1$ and the correlation length is $s_0=4.5\pm0.4$h$^{-1}$Mpc. The dashed line is the fit to our groups with the largest membership, i.e. $n_{gal}\geq 70$, with $\gamma=1.5\pm0.3$ and $s_0=11.5\pm1.1$h$^{-1}$Mpc. The fits for all the other group samples have also been estimated and appear in Table \ref{fig:table_data}. As we see our fits give $\gamma \sim 1.5-1.8$ and $s_0\sim2.0-11.5$h$^{-1}$Mpc.

We next repeat our measurements using the mock catalogues for galaxies and groups. The $\xi _{cg} (s)$ results are shown in Fig. \ref{fig:xis_mock}. The $s_0$ and $\gamma$ values from the fits on the results appear in Table \ref{fig:table_mock}. From the comparison between the $\xi _{cg} (s)$ measurements from the data and the mock catalogues we see that the results are very similar on all scales. The agreement is confirmed from the fits shown in Tables \ref{fig:table_data} and \ref{fig:table_mock}.

As already mentioned in the previous Section, for our measurements we have used Field-to-Field errors. Fig. \ref{fig:field_over_poisson2} shows how the ratio of these errors to Poisson errors, depends on the separation, s. We note that the ratio increases as we move to larger scales. The reason is that on larger scales the group-galaxy pairs become less independent causing an underestimation by Poisson errors. On our scales of interest (i.e. $2-20$h$^{-1}$Mpc), the results follow a power law, i.e. $(s/0.5$h$^{-1}$Mpc)$^{1.1}$.

\begin{figure*}
\begin{center}
\centerline{\epsfxsize = 9cm
\epsfbox{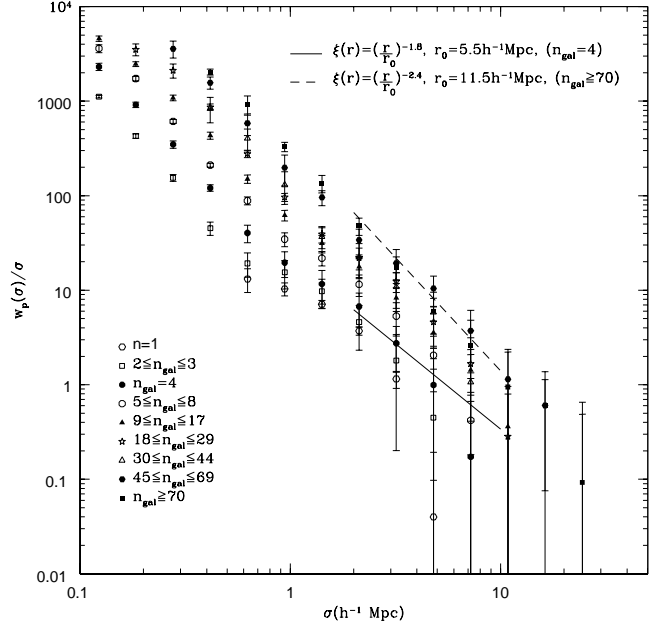}}
\caption{The galaxy group-galaxy semi-projected cross-correlation function $\omega _p(\sigma)$ for different group membership, using the data catalogues. The solid line shows the fit for the galaxy sample with $n_{gal}=4$ which gives a correlation length of $r_0=4.5$h$^{-1}$Mpc with slope of $\gamma= 2.6$. The dashed line is our fit for the galaxy groups with $n_{gal}\geq 70$ which gives $r_0=11.5$h$^{-1}$Mpc and $\gamma=2.4$.}
\label{fig:wp}
\end{center}
\end{figure*}

\begin{figure*}
\begin{center}
\centerline{\epsfxsize = 9cm
\epsfbox{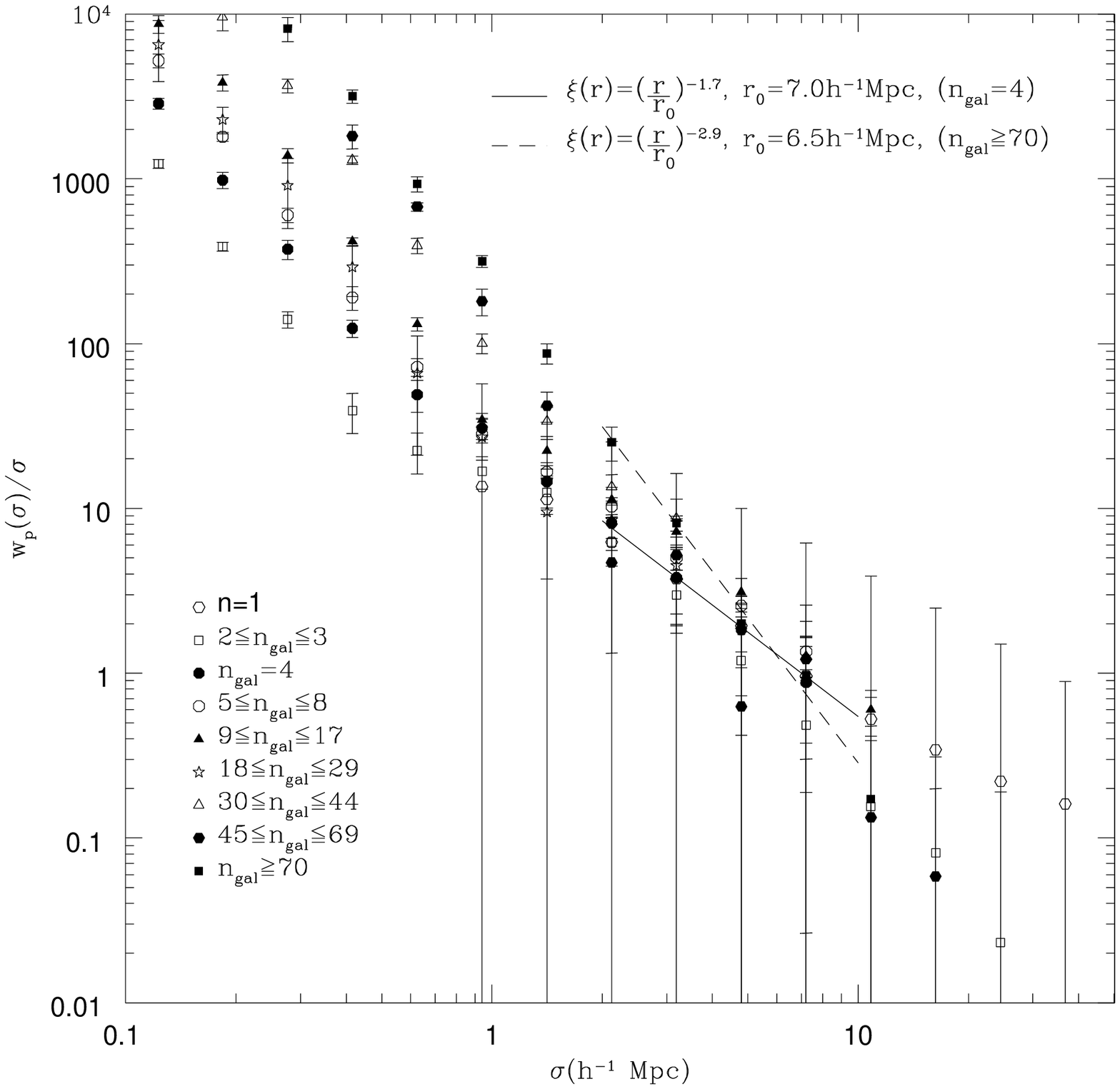}}
\caption{The galaxy group-galaxy semi-projected cross-correlation function $\omega _p(\sigma)$ for different group membership using the mock catalogues. The solid line in shows the fit for the galaxy sample with $n_{gal}=4$ which gives a correlation length of $r_0=6.5$h$^{-1}$Mpc with slope of $\gamma= 1.9$. The dashed line is our fit for the galaxy groups with $n_{gal}\geq 70$ which gives $r_0=6.5$h$^{-1}$Mpc and  $\gamma=2.9$.}
\label{fig:wp_mock}
\end{center}
\end{figure*}

\begin{figure*}
\begin{center}
\centerline{\epsfxsize = 7.5cm
\epsfbox{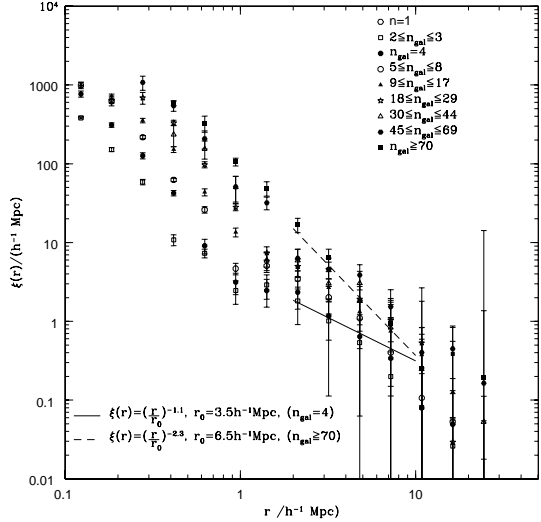}}
\caption{The real-space cross-correlation function, $\xi _{cg}(r)$, results using the data sets.}
\label{fig:xir}
\end{center}
\end{figure*}

\begin{figure*}
\begin{center}
\centerline{\epsfxsize = 7.5cm
\epsfbox{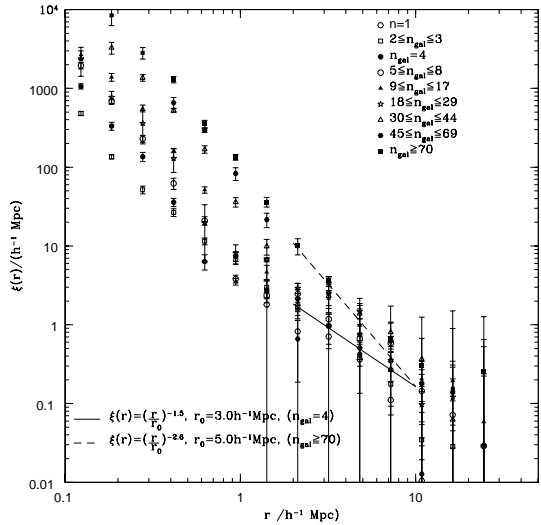}}
\caption{The real-space correlation function, $\xi _{cg}(r)$, results using the mock catalogues.}
\label{fig:xir_mock}
\end{center}
\end{figure*}

\begin{table*}
\caption{$r_0$ (or $s_0$) and $\gamma$ values for the three cross-correlation functions, i.e. $\xi _{cg}(s)$, $w_p(\sigma)/\sigma$ and $\xi _{cg}(r)$, using the data catalogues.}
\centering
\setlength{\tabcolsep}{4.0mm}
\begin{tabular}{lcccccc}
       \hline
$n_{gal}$ & \multicolumn{2}{c}{$\xi (s)$} & \multicolumn{2}{c}{$w_p(\sigma)/\sigma$}&\multicolumn{2}{c}{$\xi (r)$}\\
       \hline
$$ & $s_0$& $\gamma$ & $r_0$& $\gamma$& $r_0$& $\gamma$ \\
       \hline \hline
=1  & $1.8\pm0.2$ & $1.6\pm0.2$& $3.0\pm0.2$ & $2.7\pm0.2$  & $3.5\pm0.5$ & $1.3\pm0.2$ \\
       \hline
2-3 & $4.0\pm0.4$  & $1.5\pm0.1$ & $4.0\pm0.5$ & $2.2\pm0.2$ & $3.5\pm0.5$ & $1.2\pm0.3$  \\
       \hline
=4 & $4.5\pm0.4$  & $1.6\pm0.1$ & $5.5\pm0.5$ & $1.8\pm0.3$ & $3.5\pm0.4$ & $1.1\pm0.3$  \\
       \hline
5-8 & $5.5\pm0.4$  & $1.6\pm0.2$ & $8.0\pm0.8$ & $1.8\pm0.2$ & $5.5\pm1.0$ & $1.5\pm0.2$ \\
       \hline
9-17  & $7.0\pm0.5$  & $1.5\pm0.2$ & $10.5\pm0.9$ & $1.7\pm0.2$ & $6.5\pm1.0$ & $1.5\pm0.2$ \\
       \hline
18-29 & $7.5\pm0.6$  & $1.7\pm0.2$ & $14.0\pm1.4$ & $1.6\pm0.2$ & $9.0\pm1.0$ & $1.3\pm0.2$  \\
       \hline 
30-44 & $7.5\pm0.8$  & $1.8\pm0.2$ & $16.0\pm1.5$ & $1.5\pm0.2$ & $8.5\pm0.8$ & $1.3\pm0.2$ \\
       \hline 
45-69 & $11.0\pm0.9$  & $1.5\pm0.3$ & $15.5\pm2.0$ & $1.9\pm0.2$ & $5.0\pm1.0$ & $2.3\pm0.4$  \\
       \hline 
$\geq70$ & $11.5\pm1.1$ & $1.5\pm0.3$ & $11.5\pm1.5$ & $2.4\pm0.2$ & $6.5\pm0.6$ & $2.3\pm0.4$  \\
       \hline
\label{fig:table_data}
\end{tabular}
\end{table*}

\begin{table*}
\caption{$r_0$ (or $s_0$) and $\gamma$ values for the three cross-correlation functions, i.e. $\xi _{cg}(s)$, $w_p(\sigma)/\sigma$ and $\xi _{cg}(r)$, using the mock catalogues.}
\centering
\setlength{\tabcolsep}{4.0mm}
\begin{tabular}{lcccccc}
       \hline
$n_{gal}$ & \multicolumn{2}{c}{$\xi (s)$} & \multicolumn{2}{c}{$w_p(\sigma)/\sigma$}&\multicolumn{2}{c}{$\xi (r)$}\\
       \hline
$$ & $s_0$& $\gamma$ & $r_0$& $\gamma$& $r_0$& $\gamma$ \\
       \hline \hline
=1  & $1.8\pm0.2$ & $1.8\pm0.1$& $8.0\pm0.5$ & $1.4\pm0.1$  & $2.0\pm0.2$ & $1.2\pm0.2$ \\
       \hline
2-3 & $4.0\pm0.3$  & $1.5\pm0.1$ & $5.5\pm0.4$ & $1.9\pm0.1$ & $3.0\pm0.2$ & $1.5\pm0.2$  \\
       \hline
=4 & $4.8\pm0.4$  & $1.6\pm0.2$ & $7.0\pm0.5$ & $1.7\pm0.1$ & $3.0\pm0.2$ & $1.5\pm0.2$  \\
       \hline
5-8 & $5.0\pm0.4$  & $1.6\pm0.2$ & $9.0\pm0.9$ & $1.6\pm0.2$ & $3.5\pm0.4$ & $1.5\pm0.2$ \\
       \hline
9-17  & $6.0\pm0.4$  & $1.6\pm0.2$ & $10.5\pm1.0$ & $1.6\pm0.1$ & $5.5\pm1.0$ & $1.3\pm0.2$ \\
       \hline
18-29 & $6.0\pm0.5$  & $1.9\pm0.2$ & $7.5\pm0.8$ & $1.9\pm0.2$ & $4.5\pm0.5$ & $1.7\pm0.2$  \\
       \hline 
30-44 & $7.0\pm0.6$  & $1.8\pm0.2$ & $10.5\pm1.0$ & $1.7\pm0.2$ & $3.0\pm0.3$ & $2.2\pm0.3$ \\
       \hline 
45-69 & $7.5\pm0.7$  & $2.0\pm0.3$ & $5.0\pm0.8$ & $2.0\pm0.3$ & $4.5\pm0.6$ & $2.4\pm0.4$  \\
       \hline 
$\geq70$ & $9.0\pm1.0$ & $1.9\pm0.2$ & $6.5\pm0.5$ & $2.9\pm0.2$ & $5.0\pm0.5$ & $2.6\pm0.4$  \\
       \hline
\label{fig:table_mock}
\end{tabular}
\end{table*}

\section{The semi-projected cross-correlation function}

If $s_1$ and $s_2$ are the distances of two objects 1, 2, measured in redshift-space, and $\theta$ the angular separation between them, then $\sigma$ and $\pi$ are defined as

\begin{equation}
\pi=(s_2-s_1), $ along the line-of-sight$
\end{equation}

\begin{equation}
\sigma=\frac{(s_2+s_1)}{2}\theta , $ across the line-of-sight$
\end{equation}
The effects of redshift distortion now appear only in the radial component, $\pi$, so by integrating along the $\pi$ direction we calculate what is called the semi-projected correlation function, $\omega _p(\sigma)$

\begin{equation}
w_p(\sigma)=2\int_0^\infty \xi _{cg}(\sigma,\pi)d\pi
\end{equation}
In our case we take the upper limit of the integration to be equal to $\pi_{max}=70$h$^{-1}$Mpc. At this limit, the effect of small-scale peculiar velocities and redshift errors should be negligible. If we include very large scales, the signal will become dominated by noise and, on the other hand, if we restrict our measurements to very small scales then the amplitude will be underestimated. Now, since $\omega _p(\sigma)$ describes the real-space clustering, the last equation can be written in terms of the real-space correlation function, $\xi _{cg}(r)$, (Davis \& Peebles, 1983), i.e.

\begin{equation}
w_p(\sigma)=2\int_\sigma^{\pi_{max}}\frac{r\xi _{cg} (r)}{\sqrt(r^2-\sigma ^2)}dr
\end{equation}

Fig. \ref{fig:wp} shows the $w_p(\sigma)/\sigma$ results for the different galaxy-group samples. Since the measurements are noisier than those for $\xi (s)$, we make fits on scales of 2$\leq r\leq 10$h$^{-1}$Mpc. The solid line (Fig. \ref{fig:wp}) shows the fit for the galaxy sample with $n_{gal}=4$ which gives a correlation length of $r_0=5.5$h$^{-1}$Mpc with slope of $\gamma=1.8$. The dashed line is our fit for the galaxy groups with $n_{gal}\geq 70$ which gives $r_0=11.5$h$^{-1}$Mpc and  $\gamma=2.4$. The fits for all the other group samples appear in Table \ref{fig:table_data}. With the exception of very small groups and very rich clusters, the slope remains roughly the same, whereas the amplitude, as expected, increases with increased group membership. Fig. \ref{fig:wp_mock} shows the $w_p(\sigma)/\sigma$ results for the different galaxy-group samples using the mock catalogues and Table \ref{fig:table_mock} shows the fits (2$\leq r\leq 10$h$^{-1}$Mpc). From the comparison between the data and the mock catalogues, we see that the results are very similar on all scales.

\section{The real-space cross-correlation function}

Using the results from the semi-projected cross-correlation function, $\omega _p(\sigma)/\sigma$ and following Saunders et al. 1992, we can calculate the real-space cross-correlation function, $\xi _{cg} (r)$, as follows:

\begin{equation}
\xi _{cg} (r)=-\frac{1}{\pi}\int_r^\infty\frac{d\omega (\sigma)/d\sigma}{\sqrt{(\sigma ^2-r^2)}}d\sigma
\end{equation}
and assuming a step function for $w_p(\sigma)=w_i$ we finally get,

\begin{equation}
\xi _{cg}(\sigma _i)=-\frac{1}{\pi}\sum_{j\geq i}\frac{\omega _{j+1}-\omega _j}{\sigma _{j+1}- \sigma _j}ln{(\frac{\sigma _{j+1}+\sqrt{\sigma_ {j+1}^2-\sigma _i^2}}{\sigma _j+\sqrt{\sigma_ j^2-\sigma _i^2}})}
\end{equation}
for $r=\sigma _i$. 

The real-space cross-correlation function results are shown in Fig. \ref{fig:xir}. As in the previous Sections, Table \ref{fig:table_data} shows the $r_0$ and $\gamma$ values from the fits ($2\leq r\leq 10$h$^{-1}$Mpc). Once again, we notice that the real space cross-correlation function has a higher correlation length for groups with larger membership. The results seem to be in agreement with the $\xi _{cg}(s)$ and the $w_p(\sigma)/\sigma$ measurements, although they are much noisier. 

Then we repeat the $\xi _{cg} (r)$ measurements using the mock catalogues. The results are shown in Fig. \ref{fig:xir_mock}. The results from the fits appear in Table \ref{fig:table_mock}. From the comparison between the data and the mock catalogues, we see that the results are very similar on all scales.

\begin{figure}
\begin{center}
\centerline{\epsfxsize = 6.5cm
\epsfbox{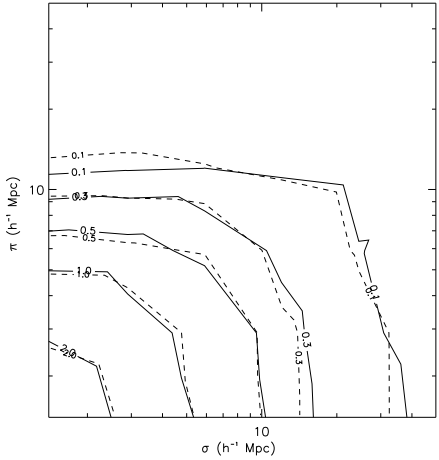}}
\caption{$\xi _{cg}(\sigma, \pi)$ results for group-galaxy sample with $n_{gal}=4$. Solid lines present the results using the data and dashed lines using the model. We see that, there is a good agreement between the two, although the data gives slightly flatter results than the model.}
\label{fig:model_vs_data_1}
\end{center}
\end{figure}

\begin{figure}
\begin{center}
\centerline{\epsfxsize = 6.5cm
\epsfbox{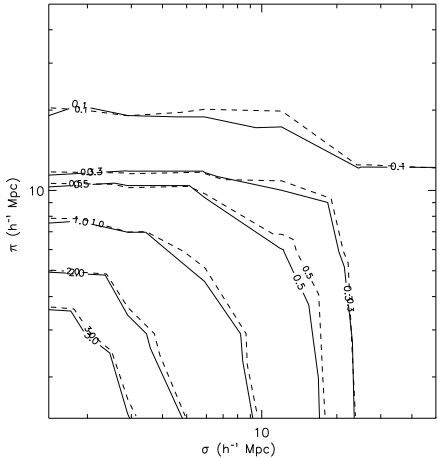}}
\caption{$\xi _{cg}(\sigma, \pi)$ results for group-galaxy sample with $9\leq n_{gal}\leq 17$. Solid lines present the results using the data and dashed lines using the model. The results agree very well, although there are some indications that at small scales the model gives slightly smaller velocity dispersions for this group-galaxy sample.}
\label{fig:model_vs_data_2}
\end{center}
\end{figure}

\begin{figure}
\begin{center}
\centerline{\epsfxsize = 6.5cm
\epsfbox{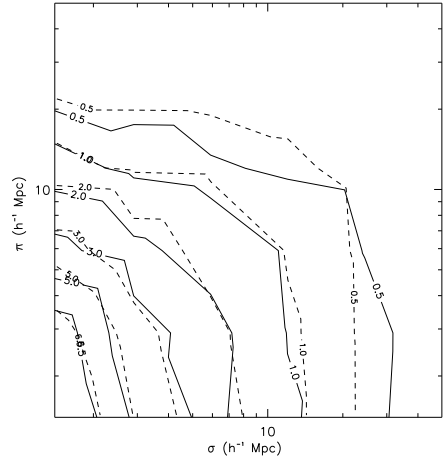}}
\caption{$\xi _{cg}(\sigma, \pi)$ results for group-galaxy sample with $45\leq n_{gal}\leq 69$. Solid lines present the results using the real data and dashed lines using the model. The two are in very good agreement.}
\label{fig:model_vs_data_3}
\end{center}
\end{figure}

\section{Constraining $\beta$ from redshift-space distortions}

\subsection{The $\xi _{cg}(\sigma, \pi)$  cross-correlation function}

Now we shall present our $\xi _{cg}(\sigma, \pi)$ results from the group-galaxy data samples as well as the mock catalogues. These results will be used later in order to model the redshift-space distortions. More precisely, the shape of the $\xi _{cg}(\sigma, \pi)$ measurements will be used.             
            
Figures \ref{fig:model_vs_data_1}-\ref{fig:model_vs_data_3} show our results for galaxy groups with $n_{gal}=4$, $9\leq n_{gal}\leq 17$ and $45\leq n_{gal}\leq 69$ (solid lines). Comparing the results from different samples we see that, moving to bigger galaxy groups, differentiations from the nearly symmetric clustering pattern of $n_{gal}=4$ group-galaxies start to increase. Elongations in the redshift direction along the line-of-sight ($\pi$-direction) are most evident for the larger membership galaxy-group samples of $45\leq n_{gal}\leq 69$.

\subsection{Model description}

We assume that the effects of redshift-space distortions can be modelled by adjusting $\xi (r)$ for the effects of velocity dispersion and a simple infall model. The model we assume for the bias is, 

\begin{equation}
\xi_{cm}=\frac{\xi _{cg}}{b}
\end{equation}
Next we introduce the infall velocity of the galaxies, $\upsilon (r_z)$, as a function of the real-space separation along the $\pi$ direction, $r_z$, adapting eqn 77.24 of Peebles (1980) and assuming a power law of slope $-\gamma$ for $\xi _{cg}(r)$ we get, 

\begin{equation}
\upsilon (r_z)= -\frac{1}{3-\gamma}\beta H(z)r_z\xi _{cg}(r),
\label{eqn:model}
\end{equation} 
where we have substituted $\beta=\frac{\Omega ^{0.6}}{b}$. Equation \ref{eqn:model} applies in the linear regime ($\xi _{cm}\lesssim1$) for infall into rich clusters. We accept it is an approximation to apply equation \ref{eqn:model} to groups, where group-group interactions may be non-negligible.

The magnitude of the elongation along the $\pi$-direction of the $\xi _{cg}(\sigma, \pi )$ plot caused by the galaxy velocity dispersion is denoted by $<\omega _z^2>^{1/2}$, which can be expressed by a Gaussian (Ratcliffe et al. 1996), as

\begin{equation}
f(\omega _z)=\frac{1}{\sqrt{2\pi}<\omega_z^2>^{1/2}}exp(-\frac{1}{2}\frac{|\omega_z|^2}{<\omega_z^2>^{1/2}})
\label{eqn:f_vel_dis}
\end{equation}
In order to include the small scale redshift-space effects due to the random motions of galaxies, we convolve the $\xi _{cg} (\sigma, \pi)$ model with the velocity dispersion distribution, given by the equation \ref{eqn:f_vel_dis}. Then $\xi _{cg} (\sigma, \pi)$ is given by (Peebles 1980, Hoyle 2000) 

\begin{equation}
1+\xi (\sigma, \pi)=\int_{-\infty}^{+\infty}{(1+\xi (r))f(\omega _z)d\omega _z}
\label{eqn:sigma2}
\end{equation}
and modifying this to include the effects of the bulk motions we finally get 

\begin{equation}
1+\xi _{cg}(\sigma, \pi)=\int_{-\infty}^{+\infty}{(1+\xi _{cg}(r))f(\omega _z-\upsilon (r_z)))d\omega _z}
\label{eqn:sigma3}
\end{equation}
This is the $\xi _{cg}(\sigma ,\pi )$ we get from our model and following Hoyle et al. 2002 fitting procedure we put constraints on $\beta _{cg}$ for each one of our group-galaxy samples.

\begin{table*}
\caption{Estimation of $<w_z^2>^{1/2}$ for different galaxy group memberships.}
\centering
\setlength{\tabcolsep}{3.0mm}
\begin{tabular}{lcccccc}
       \hline
$ $ & \multicolumn{5}{c}{$<w_z^2>^{1/2} (km/s)$} & $ $ \\
       \hline
$ $ & \multicolumn{3}{c}{data} & \multicolumn{3}{c}{mock} \\
$$ & fixed & $\chi ^2$ & $\Omega _m^0=const$ & fixed & $\chi ^2$  & $\Omega _m^0=const$ \\
       \hline \hline
=1  & $$ & $250$ & $200_{-30}^{+10}$ & $$ & $240$& $200_{-200}^{+90}$ \\
       \hline
2-3  & $160$ & $280$  & $200_{-10}^{+10}$ & $165$ & $280$& $250_{-60}^{+60}$  \\
       \hline
=4  & $174$ & $250$  & $200_{-10}^{+30}$ & $190$ & $280$& $200_{-90}^{+10}$  \\
       \hline
5-8 & $221$  & $300$ & $300_{-40}^{+20}$ &  $226$  & $300$ & $300_{-25}^{+5}$ \\
       \hline
9-17  & $285$  & $290$ & $200_{-10}^{+30}$ &  $299$  & $280$ & $200_{-10}^{+15}$ \\
       \hline
18-29 & $371$  & $280$ & $200_{-15}^{+20}$ & $388$  & $280$ & $150_{-150}^{+10}$ \\
       \hline
30-44 & $420$  & $270$ & $250_{-250}^{+60}$ &  $443$  & $260$& $300_{-300}^{+50}$ \\
       \hline
45-69 & $488$  & $430$ & $400_{-130}^{+110}$ & $472$  & $450$& $400_{-130}^{+130}$   \\
       \hline
$\geq70$ & $606$  & $410$ & $400_{-130}^{+110}$ & $588$  & $390$& $300_{-280}^{+110}$\\
       \hline
\label{table:w1}
\end{tabular}
\end{table*}

\begin{table*}
\caption{Estimation of $\beta$ for different galaxy group memberships.}
\centering
\setlength{\tabcolsep}{4.0mm}
\begin{tabular}{lcccccc}
       \hline
$ $ & \multicolumn{3}{c}{data} & \multicolumn{3}{c}{mock} \\
       \hline 
$n_{gal}$ & $\beta (fixed)$ & $\beta (\chi^2)$& $\Omega _m^0=const$ & $\beta (fixed)$ & $\beta (\chi^2)$& $\Omega _m^0=const$ \\
       \hline \hline
=1 & $$  & $4.00_{-2.13}^{+1.60}$ & $1.50_{-1.00}^{+2.70}$ & $$  & $6.50_{-4.00}^{+1.50}$ & $4.25_{-2.55}^{+0.65}$  \\
       \hline
2-3 & $4.80_{-0.60}^{+0.40}$  & $2.60_{-0.40}^{+0.35}$ & $2.75_{-0.25}^{+0.25}$& $6.50_{-1.25}^{+2.51}$  & $3.80_{-0.80}^{+1.60}$ & $4.25_{-1.00}^{+1.55}$ \\
       \hline
=4 & $2.55_{-0.40}^{+0.50}$  & $1.80_{-0.41}^{+0.50}$ & $1.80_{-0.30}^{+0.50}$& $2.75_{-0.62}^{+0.60}$  & $2.50_{-0.90}^{+0.95}$ & $2.20_{-0.40}^{+1.55}$ \\
       \hline
5-8 & $1.65_{-0.25}^{+0.17}$  & $1.35_{-0.35}^{+0.28}$ & $1.35_{-0.35}^{+0.30}$&  $1.50_{-0.20}^{+0.29}$  & $1.25_{-0.20}^{+0.35}$ & $1.20_{-0.15}^{+0.35}$\\
       \hline
9-17  & $0.75_{-0.24}^{+0.21}$  & $0.75_{-0.23}^{+0.20}$ & $0.90_{-0.17}^{+0.08}$&  $0.60_{-0.12}^{+0.21}$  & $0.60_{-0.17}^{+0.23}$ & $0.80_{-0.10}^{+0.20}$ \\
       \hline
18-29 & $0.40_{-0.14}^{+0.07}$  & $0.30_{-0.07}^{+0.12}$ & $0.30_{-0.10}^{+0.08}$& $0.30_{-0.30}^{+0.50}$  & $0.30_{-0.28}^{+0.22}$ & $0.30_{-0.30}^{+0.30}$   \\
       \hline 
30-44 & $0.30_{-0.30}^{+0.27}$  & $0.30_{-0.30}^{+0.20}$ & $0.30_{-0.30}^{+0.12}$& $0.40_{-0.28}^{+0.25}$  & $0.50_{-0.23}^{+0.35}$ & $0.80_{-0.30}^{+0.50}$   \\
       \hline 
45-69 & $1.20_{-0.48}^{+0.55}$  & $1.30_{-0.50}^{+0.45}$ & $1.20_{-0.35}^{+0.25}$& $1.50_{-1.20}^{+3.20}$  & $1.50_{-1.20}^{+2.95}$ & $1.20_{-0.45}^{+0.35}$  \\
       \hline 
$\geq70$ & $1.50_{-0.52}^{+0.35}$  & $2.50_{-1.10}^{+0.71}$ & $2.50_{-0.80}^{+0.30}$& $1.80_{-0.45}^{+0.21}$  & $1.90_{-0.43}^{+0.11}$ & $2.50_{-0.90}^{+3.00}$  \\
       \hline
\label{table:beta1}
\end{tabular}
\end{table*}

\begin{figure*}
\hfill
  \begin{minipage}{.45\textwidth}
\begin{center}
\centerline{\epsfxsize = 5.0cm
\epsfbox{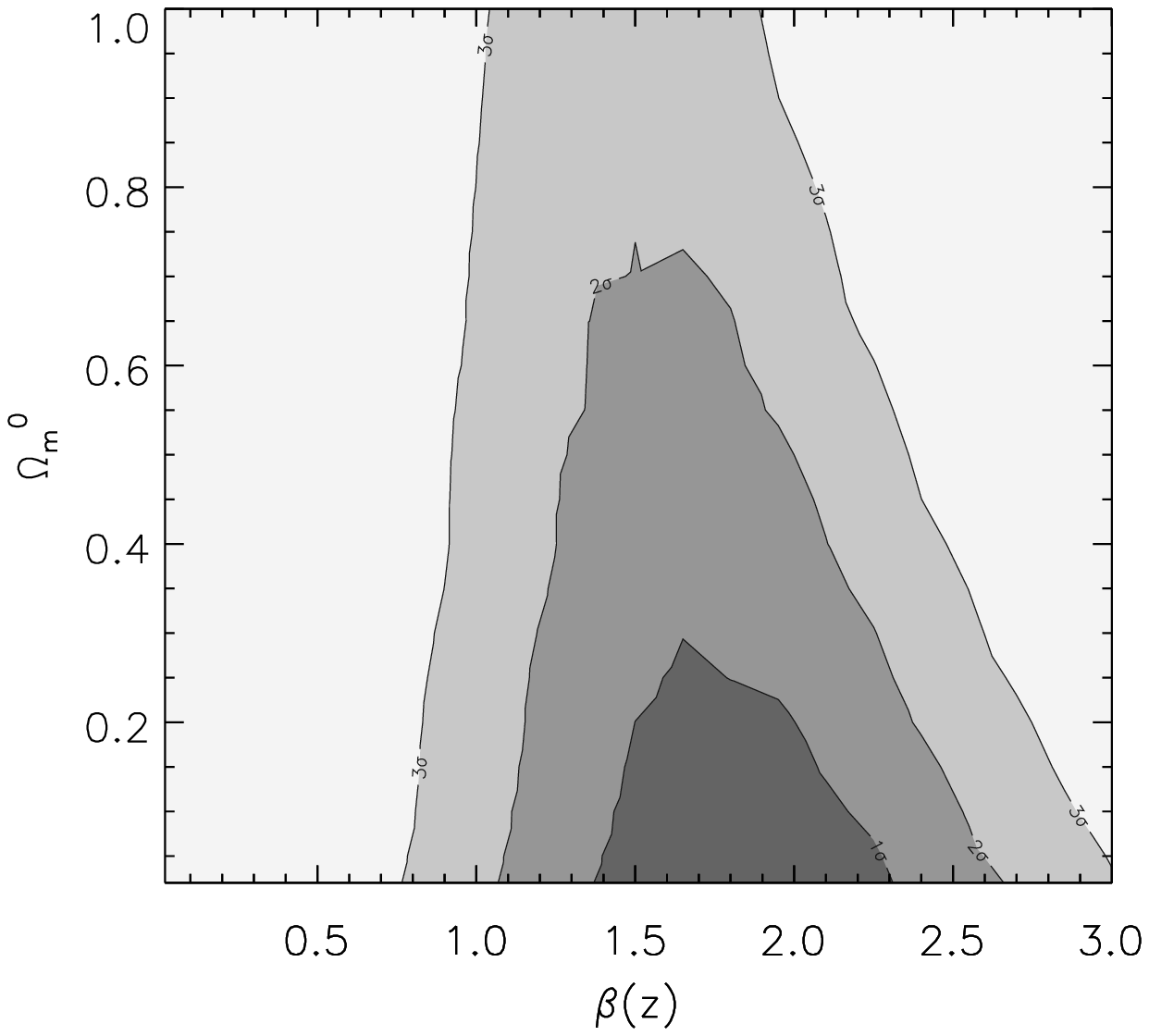}}
\caption{Likelihood contours of $\Omega_m^0-\beta (z=0.11)$ for the $n_{gal}=4$ data group-galaxy sample. A $\Lambda$CDM cosmology is assumed along with a model where  $r_0=4.5$h$^{-1}$Mpc and $\gamma =1.6$ and velocity dispersion $<w_z^2>^{1/2}=250$kmsec$^{-1}$. The best fit value for $\beta$ is $\beta=1.80_{-0.41}^{+0.50}$.}
\label{fig:omega4_2}
\end{center}
\end{minipage}
\hfill
\begin{minipage}{.45\textwidth}
\begin{center}
\centerline{\epsfxsize = 5.0cm
\epsfbox{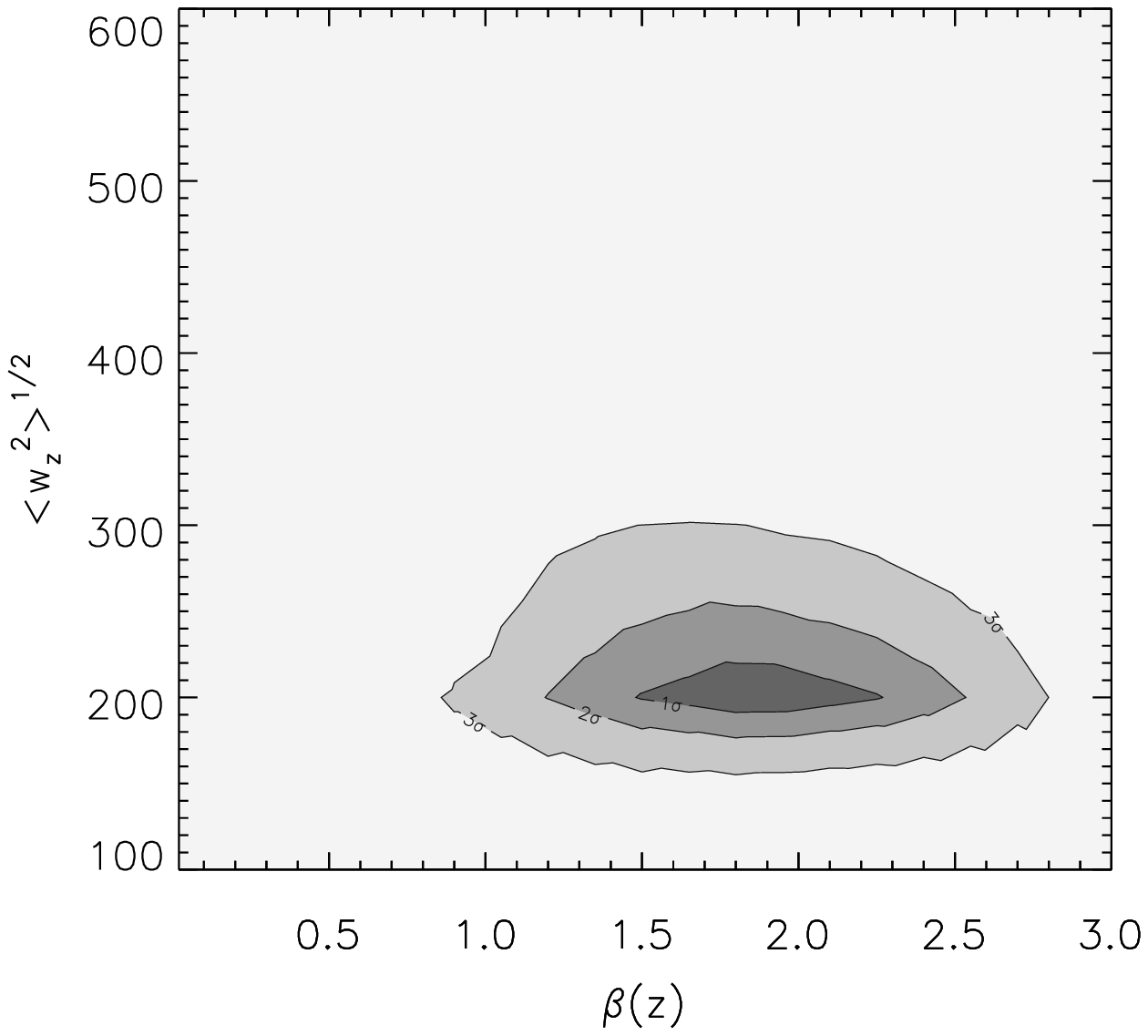}}
\caption{Likelihood contours of $<w_z^2>^{1/2}-\beta (z=0.11)$ for the $n_{gal}=4$ data group-galaxy sample. A $\Lambda$CDM cosmology is assumed along with a model where  $r_0=4.5$h$^{-1}$Mpc and $\gamma =1.6$ and $\Omega_m^0=0.3$. The best fit values are $\beta=1.80_{-0.30}^{+0.50}$ and $<w_z^2>^{1/2}=200_{-10}^{+30}$.}
\label{fig:w4_2}
\end{center}
\end{minipage}
  \hfill
\end{figure*}

\begin{figure*}
\hfill
  \begin{minipage}{.45\textwidth}
\begin{center}
\centerline{\epsfxsize = 5.0cm
\epsfbox{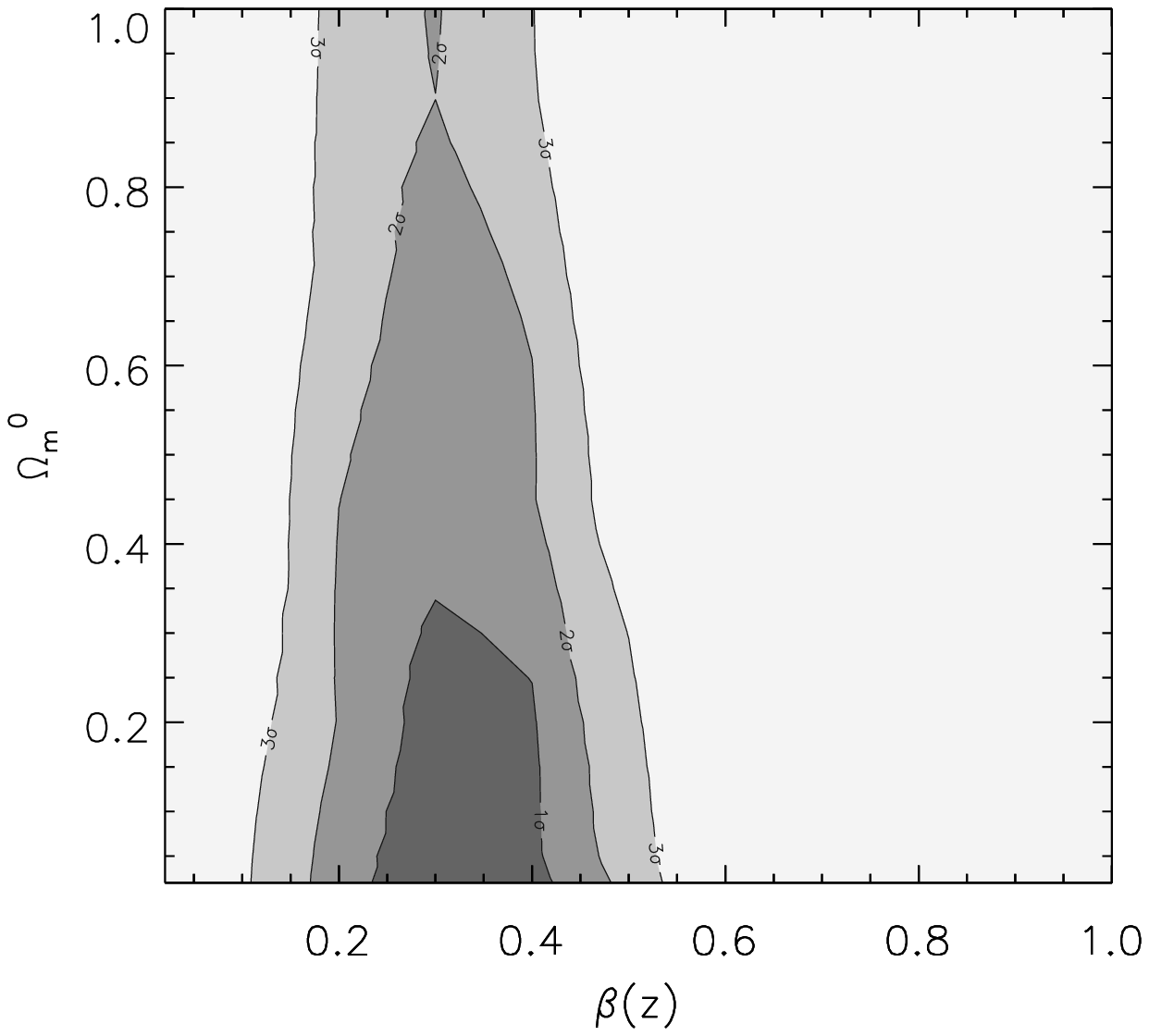}}
\caption{Likelihood contours of $\Omega_m^0-\beta (z=0.11)$ using model for the $18\leq n_{gal}\leq 29$ data group-galaxy sample. A $\Lambda$CDM cosmology is assumed along with a model where $r_0=7.5$, $\gamma=1.7$ with $<w_z^2>^{1/2}=280$kmsec$^{-1}$. The best fit value for $\beta$ is $\beta=0.30_{-0.07}^{+0.12}$.} 
\label{fig:omega18_2}
\end{center}
\end{minipage}
\hfill
\begin{minipage}{.45\textwidth}
\begin{center}
\centerline{\epsfxsize = 5.0cm
\epsfbox{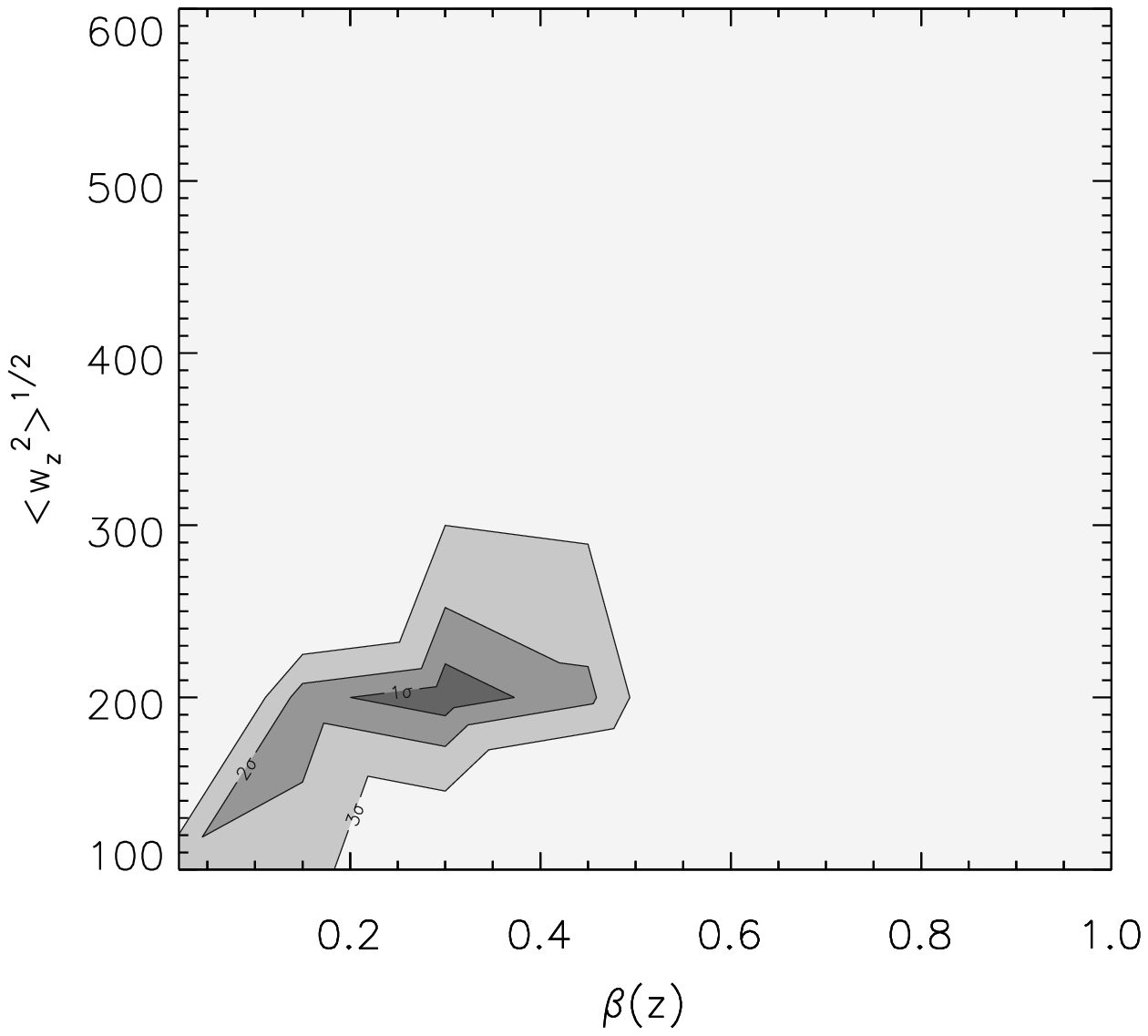}}
\caption{Likelihood contours of $<w_z^2>^{1/2}-\beta (z=0.11)$ for the $18\leq n_{gal}\leq 29$ data group-galaxy sample. A $\Lambda$CDM cosmology is assumed along with a model where $r_0=7.5$, $\gamma=1.7$ and $\Omega_m^0=0.3$. The best fit values are $\beta=0.30_{-0.10}^{+0.08}$ and $<w_z^2>^{1/2}=200_{-15}^{+20}$.} 
\label{fig:w18_2}
\end{center}
\end{minipage}
  \hfill
\end{figure*}

\begin{figure*}
\hfill

  \begin{minipage}{.45\textwidth}
\begin{center}
\centerline{\epsfxsize = 5.0cm
\epsfbox{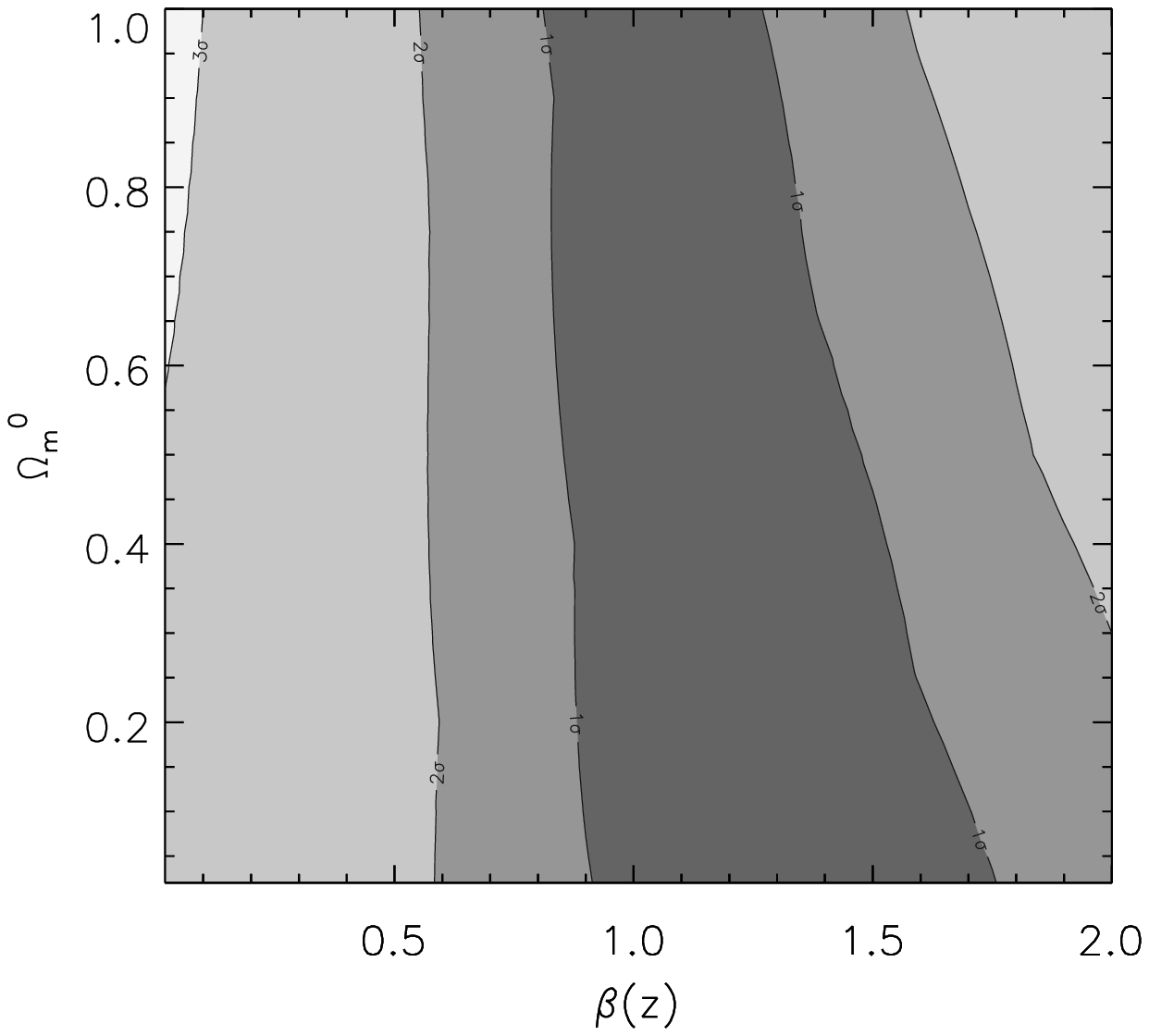}}
\caption{Likelihood contours of $\Omega_m^0-\beta (z=0.11)$ using model for the $45\leq n_{gal}\leq 69$ group sample. A $\Lambda$CDM cosmology is assumed along with a model where $r_0=11.0$, $\gamma=1.5$ with $<w_z^2>^{1/2}=430$kmsec$^{-1}$. The best fit value for $\beta$ is $\beta=1.30_{-0.50}^{+0.45}$.} 
\label{fig:omega45_2}
\end{center}
\end{minipage}
\hfill
\begin{minipage}{.45\textwidth}
\begin{center}
\centerline{\epsfxsize = 5.0cm
\epsfbox{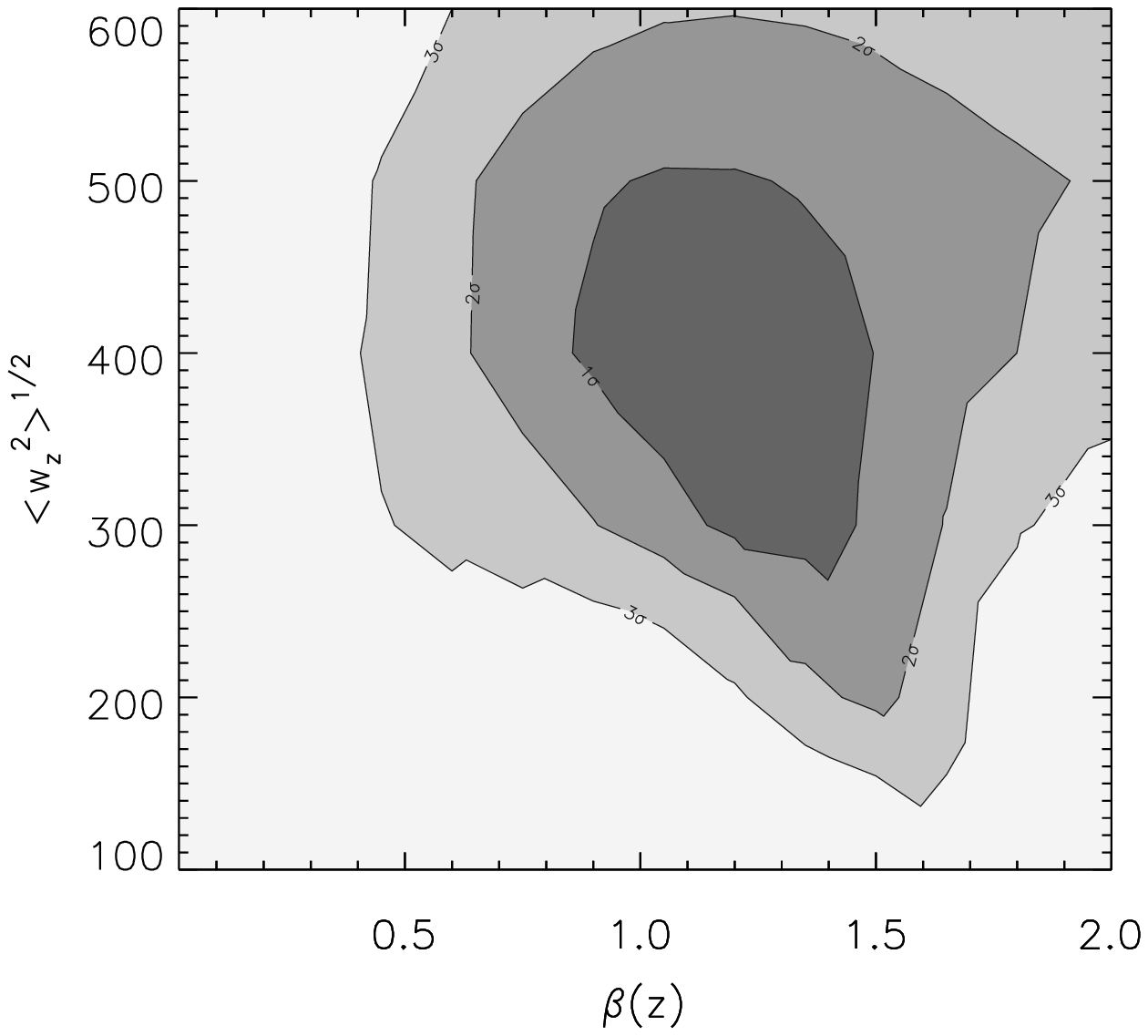}}
\caption{Likelihood contours of $<w_z^2>^{1/2}-\beta (z=0.11)$ for the $45\leq n_{gal}\leq 69$ group sample. A $\Lambda$CDM cosmology is assumed along with a model where $r_0=11.0$, $\gamma=1.5$ and $\Omega_m^0=0.3$. The best fit values are $\beta=1.20_{-0.35}^{+0.30}$ and $<w_z^2>^{1/2}=400_{-130}^{+110}$.} 
\label{fig:w45_2}
\end{center}
\end{minipage}
  \hfill
\end{figure*}

\begin{figure*}
\hfill
  \begin{minipage}{.45\textwidth}
\begin{center}
\centerline{\epsfxsize = 8.5cm
\epsfbox{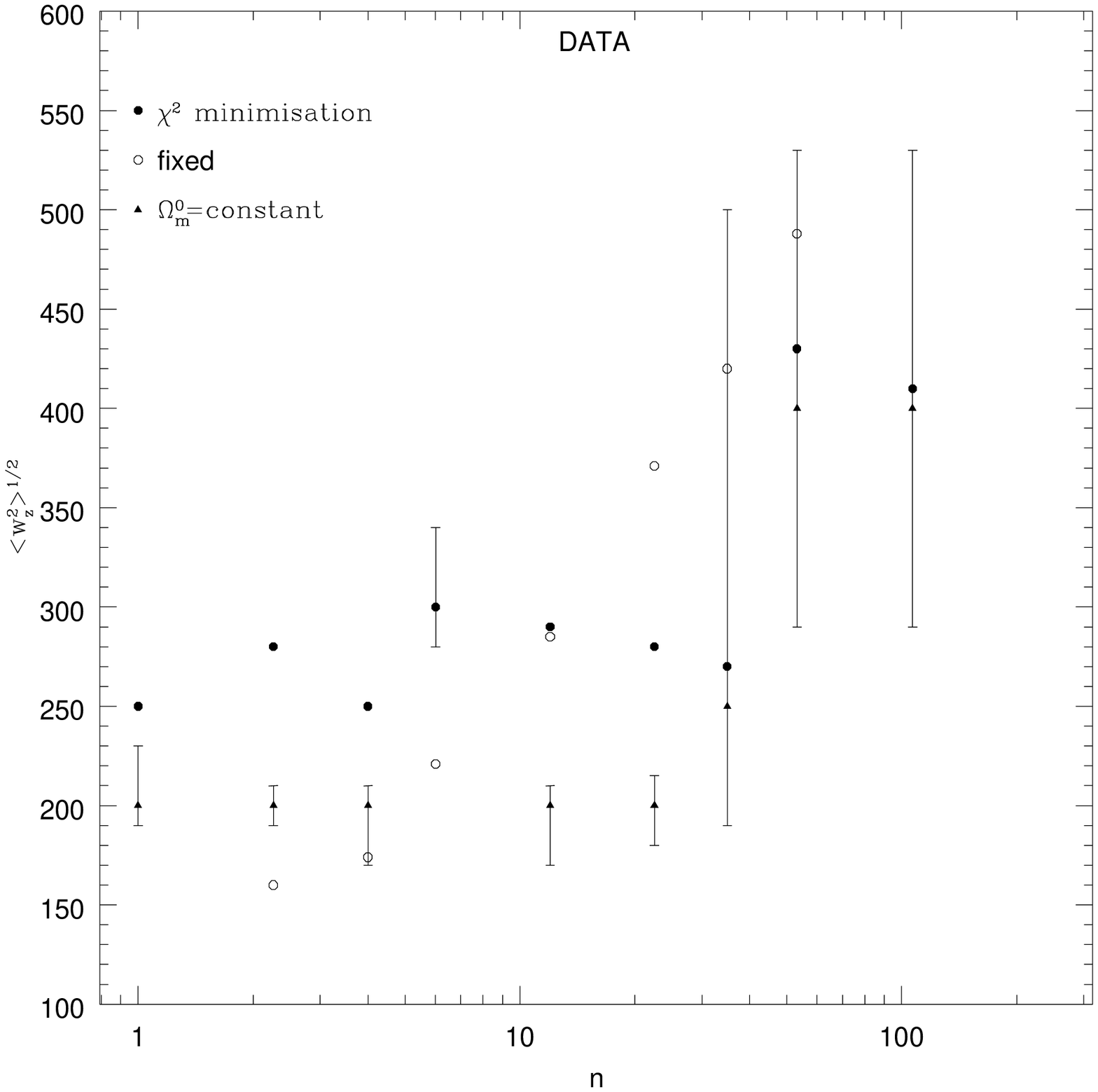}}
\caption{$<w_z^2>^{1/2}$ vs membership for the data. Filled circles show the results when using $\chi ^2$ minimisation to estimate $<w_z^2>^{1/2}$, open circles show the fixed values for $<w_z^2>^{1/2}$ and triangles when we set $\Omega _m^0=0.3$ and let $<w_z^2>^{1/2}$ vary.}
\label{fig:w_data}
\end{center}
\end{minipage}
\hfill
\begin{minipage}{.45\textwidth}
\begin{center}
\centerline{\epsfxsize = 8.5cm
\epsfbox{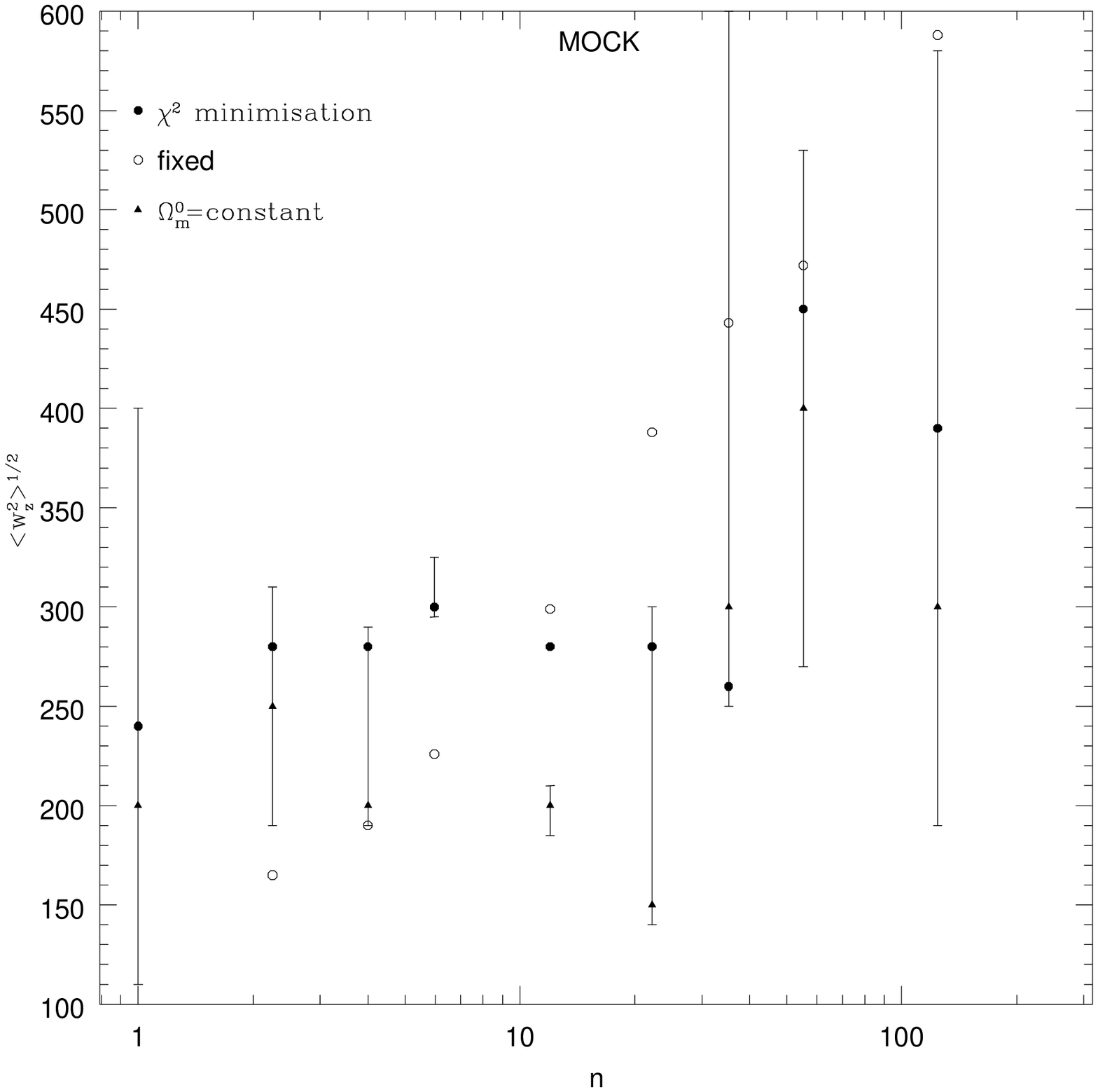}}
\caption{$<w_z^2>^{1/2}$ vs membership for the mock. Filled circles show the results when using $\chi ^2$ minimisation to estimate $<w_z^2>^{1/2}$, open circles show the fixed values for $<w_z^2>^{1/2}$ and triangles when we set $\Omega _m^0=0.3$ and let $<w_z^2>^{1/2}$ vary.}
\label{fig:w_mock}
\end{center}
\end{minipage}
  \hfill
\end{figure*}

\begin{figure*}
\begin{center}
\centerline{\epsfxsize = 8.5cm
\epsfbox{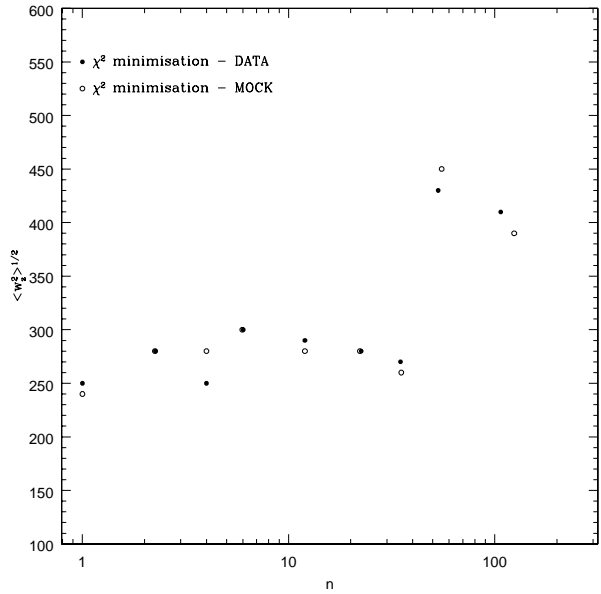}}
\caption{$<w_z^2>^{1/2}$ vs membership, a comparison between the data (filled circles) and the mocks (open circles) using  $\chi ^2$ minimisation to estimate $<w_z^2>^{1/2}$. We notice the jump of the velocity dispersion for the group samples with the two largest memberships.}
\label{fig:w_data_mock}
\end{center}
\end{figure*}

\begin{figure*}
\hfill
  \begin{minipage}{.45\textwidth}
\begin{center}
\centerline{\epsfxsize = 8.5cm
\epsfbox{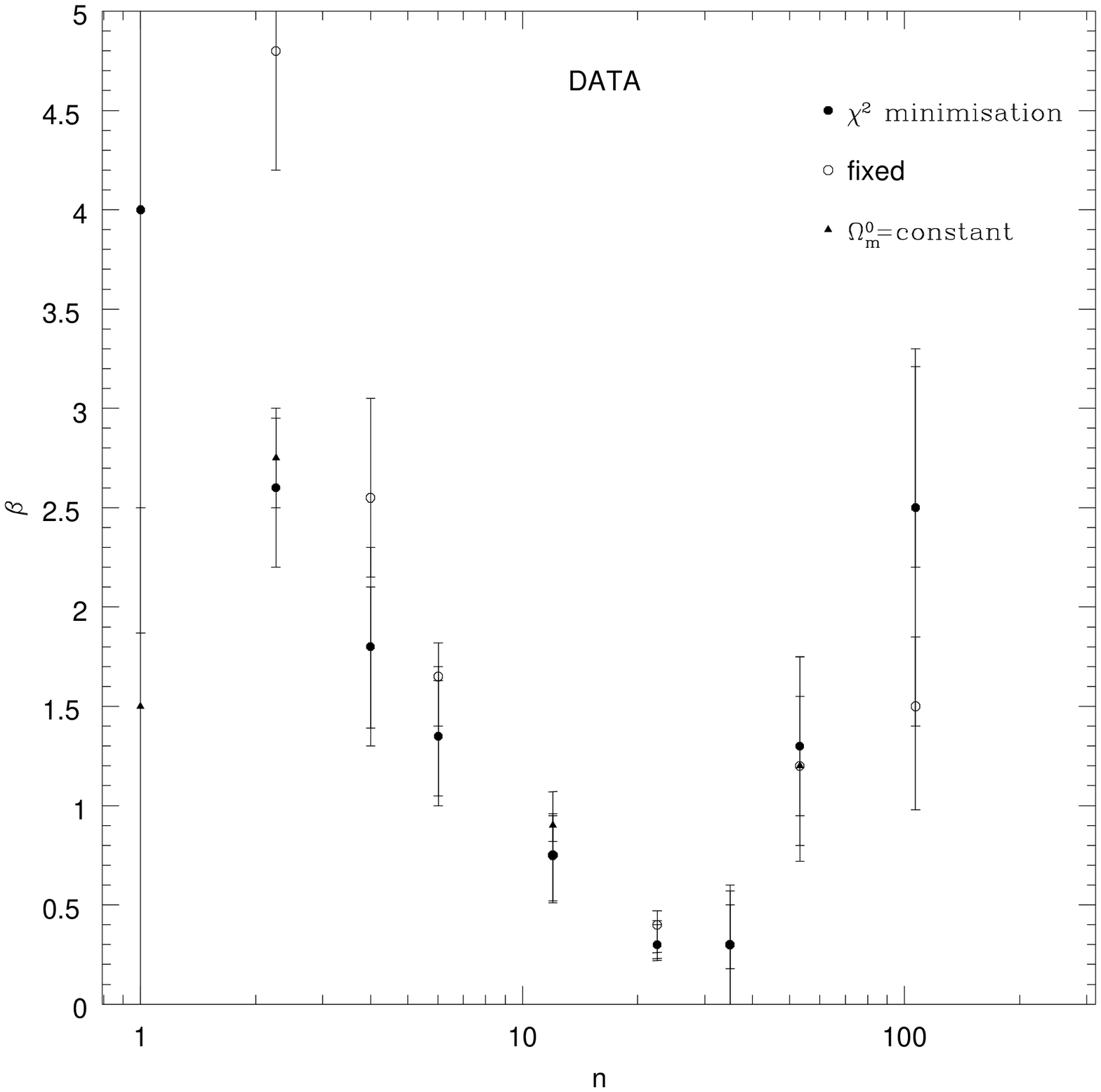}}
\caption{$\beta$ vs membership for the data. Filled circles show the results when using $\chi ^2$ minimisation to estimate $<w_z^2>^{1/2}$, open circles when we use the fixed values for $<w_z^2>^{1/2}$ and triangles when we set $\Omega _m^0=0.3$ and let $<w_z^2>^{1/2}$ vary.}
\label{fig:b_data}
\end{center}
\end{minipage}
\hfill
\begin{minipage}{.45\textwidth}
\begin{center}
\centerline{\epsfxsize = 8.5cm
\epsfbox{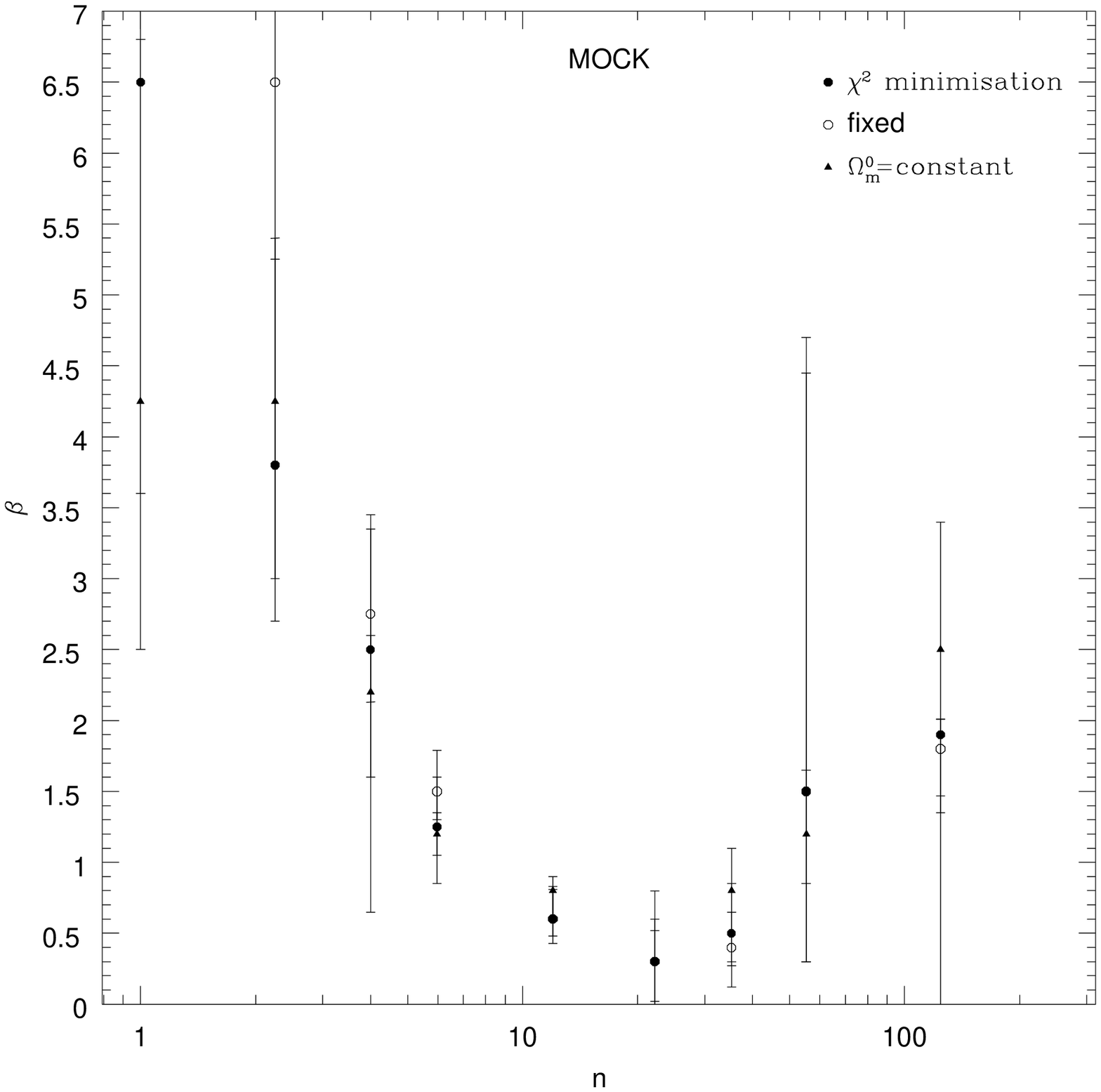}}
\caption{$\beta$ vs membership for the mock. Filled circles show the results when using $\chi ^2$ minimisation to estimate $<w_z^2>^{1/2}$, open circles when we use the fixed values for $<w_z^2>^{1/2}$ and triangles when we set $\Omega _m^0=0.3$ and let $<w_z^2>^{1/2}$ vary.}
\label{fig:b_mock}
\end{center}
\end{minipage}
  \hfill
\end{figure*}

\begin{figure*}
\begin{center}
\centerline{\epsfxsize = 8.5cm
\epsfbox{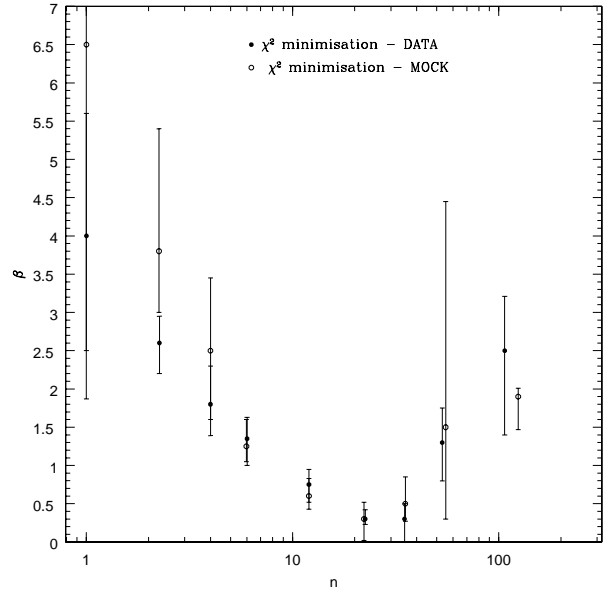}}
\caption{$\beta$ vs membership, a comparison between the data (filled circles) and the mocks (open circles) using $\chi ^2$ minimisation to estimate $<w_z^2>^{1/2}$.}
\label{fig:b_data_mock}
\end{center}
\end{figure*}

\subsection{Results from modelling redshift-space distortions}

In our analysis we have used three methods, which differ in their treatment of the galaxy velocity dispersion. For our first method, we use the average velocity dispersions of our group samples, as assigned by the 2PIGG team. We shall call these, fixed group velocity dispersions. It should be noted, that in the measurement of these $<w_z^2>^{1/2}$ the 2PIGG team has included the redshift measurement error which will contribute $\sigma _{err}\sim85$kms$^{-1}$ (Eke et al. 2004a) to their velocity dispersion measurements. 

The second method is to fit the group velocity dispersion, i.e. choose that which gives the smallest $\chi ^2$ value to our $\Omega _m^0-\beta$ estimates from $\xi _{cg}(\sigma, \pi)$. We shall call these $\chi ^2$ velocity dispersions of our samples. We do not expect the two group velocity dispersions (fixed and $\chi ^2$) to be the same because our fits to $\xi _{cg}$ take into account both the finger$-$of$-$god effect and the dynamical flattening in the $\pi$ direction. On the other hand, the 2PIGG velocity dispersion estimation ignores dynamical infall, at least before any calibration is applied from the mock catalogues. 

The third and final method is to keep $\Omega _m^0=0.3$ constant, and instead fit for the group velocity dispersion. Thus, instead of fitting $\Omega _m^0-\beta$ we fit for $<w_z^2>^{1/2}-\beta$ at fixed $\Omega _m^0$.

For all three methods, the assumed cosmology has $\Omega _m=0.3$ and $\Omega _\Lambda=0.7$. We also use the $s_0$ and $\gamma$ values from the fits on the $\xi _{cg}(s)$ measurements, shown in Tables  \ref{fig:table_data} and \ref{fig:table_mock} for the data and the mocks, respectively. Nevertheless, as already noted, we let $s_0$ vary as a free parameter. As has been pointed out, the fits on the $\xi _{cg}(s)$ measurements have been made on scales of $2-20$h$^{-1}$Mpc. However, in our fitting procedure we use scales of $5-40$h$^{-1}$Mpc. Since the errors increase at larger scales we have simply extrapolated our smaller scale fits into this region. The reason for moving further away from the group centres is that the infall model in equation \ref{eqn:model} only applies on large linear scales. Our results, both for the group-galaxy velocity dispersions, $<w_z^2>^{1/2}$, and the infall parameters, $\beta$, are shown in Tables \ref{table:w1} and \ref {table:beta1}. In Figures  \ref{fig:omega4_2}, \ref{fig:omega18_2} and \ref{fig:omega45_2} we show the $\Omega_m^0-\beta (z=0.11)$ and $<w_z^2>^{1/2}-\beta (z=0.11)$ contours (method 2 and 3) for data group-galaxy samples with $n_{gal}=4$, $18\leq n_{gal}\leq 29$ and $45\leq n_{gal}\leq 69$, respectively.

Figures \ref{fig:model_vs_data_1}-\ref{fig:model_vs_data_3} show our model results for galaxy groups with $n_{gal}=4$, $9\leq n_{gal}\leq 17$ and $45\leq n_{gal}\leq 69$ (dashed lines). The results from the fitted models (dashed lines) are consistent with those from the data (solid lines) in all cases.

\subsection{Discussion of the results}

Fig. \ref{fig:w_data} shows a comparison of the $<w_z^2>^{1/2}$ values using the three different methods, for the data. We see, that the fixed values (open circles) increase with increased group membership whereas the $\chi^2$ velocity dispersions (filled circles) stay roughly constant for small and intermediate group-galaxy samples and rise for clusters. The reason for this discrepancy may be due to the different methods, as explained above. Our third method (keep $\Omega _m^0$ constant, triangles) follows (roughly) the same pattern as the $\chi ^2$ fits. The same pattern is repeated for the mocks in Fig. \ref{fig:w_mock}. Finally, Fig. \ref{fig:w_data_mock} shows a data-mock comparison using the $\chi^2$ group velocity dispersions. The two results are in very good agreement and in both cases we notice the big jump that the velocity dispersion makes for the group samples with the two largest memberships.  
 
Fig. \ref{fig:b_data} shows a comparison of the $\beta$ values using the three different methods, for the data. We see that all three give consistent results. Small group-galaxies have a large infall parameter which decreases as we move to larger membership group-galaxy samples. Then, it rises up again for cluster-galaxies. This means that the implied bias rises by the same factor ($\approx 8\times$). The same pattern is followed using the mock catalogues, as we see in Fig. \ref{fig:b_mock}. The only difference is that the factor is now slightly higher than in the data case. Of course, in the case of mocks, the minimum in $\beta$ is expected, as the efficiency of galaxy formation is modelled to be lower at high and low halo masses. Finally, Fig. \ref{fig:b_data_mock} shows a comparison between the data and the mocks using the $\chi^2$ $<w_z^2>^{1/2}$. Once again, the agreement is very good.  

As already mentioned the infall parameter, $\beta$, estimated in this Section is a bias indicator. Therefore, we would like to compare our Figures \ref{fig:b_data} and \ref{fig:b_mock} with the M/L results of Eke et al. 2004b and 2006. Both results show a similar behaviour, i.e. the existence of a minimum value for $\beta$ (or M/L) which rises for small groups and rich clusters. What we would like to check is the agreement concerning the position of this minimum. This is the subject of the next Section.

\section{Galaxy group luminosities}

\subsection{Calculation of group luminosities} 

In this Section we shall convert our average group memberships to the corresponding luminosities in order to compare our results with Eke et al. 2004b and 2006. First we calculate the observed luminosity for each galaxy, using the apparent magnitude, m ($=b_j$), given by the SDSS team, i.e.

\begin{equation}
M_{gal/lim}=b_j^{gal/lim}-25-5logd_L-\frac{z+6z^2}{1+\frac{8}{9}z^{2.5}}
\end{equation}
where $M_{gal}$ and $M_{lim}$ is the absolute magnitude of the galaxy and limiting $b_j$ absolute magnitude of the survey at the position of this galaxy, respectively. $d_L$ is the luminosity distance, calculated by $d_L=\frac{cz}{H_0}$, which is a good approximation for our low redshift galaxies ($H_0=100$kms$^{-1}$Mpc$^{-1}$). The last term is the k-correction (Norberg et al. 2002). Then the corresponding luminosities are:

\begin{equation}
\frac{L_{gal/lim/*}}{L_{\odot}}=10^{-0.4(M_{gal/lim/*}-M_{\odot})}
\end{equation}
where $L_{\odot}$, $M_{\odot}$ are the solar luminosity and absolute magnitude and $M_*$ ($=-19.725$) and $L_*$ are the characteristic galaxy absolute magnitude and luminosity, respectively. Then the observed luminosity of a group is just the sum of the group galaxy luminosities, i.e.

\begin{equation}
\frac{L_{group}}{L_{\odot}}=\sum_i^n{\frac{L_{gal}(i)}{L_{\odot}}}
\end{equation}
We then correct this to include the contribution from galaxies that have luminosities below the luminosity limit ($L_{lim}$), at the group redshift. This is done by dividing the observed group luminosity by the incomplete $\Gamma$ function, $\frac{\Gamma(\alpha +2, \frac{L_{min}}{L_*})}{\Gamma(\alpha +2)}$, where $\alpha$ ($=-1.18$, Eke et al. 2004b) is the power law index for the faint end slope (Schechter 1976).

\begin{figure*}
\hfill
  \begin{minipage}{.45\textwidth}
\begin{center}
\centerline{\epsfxsize = 9.0cm
\epsfbox{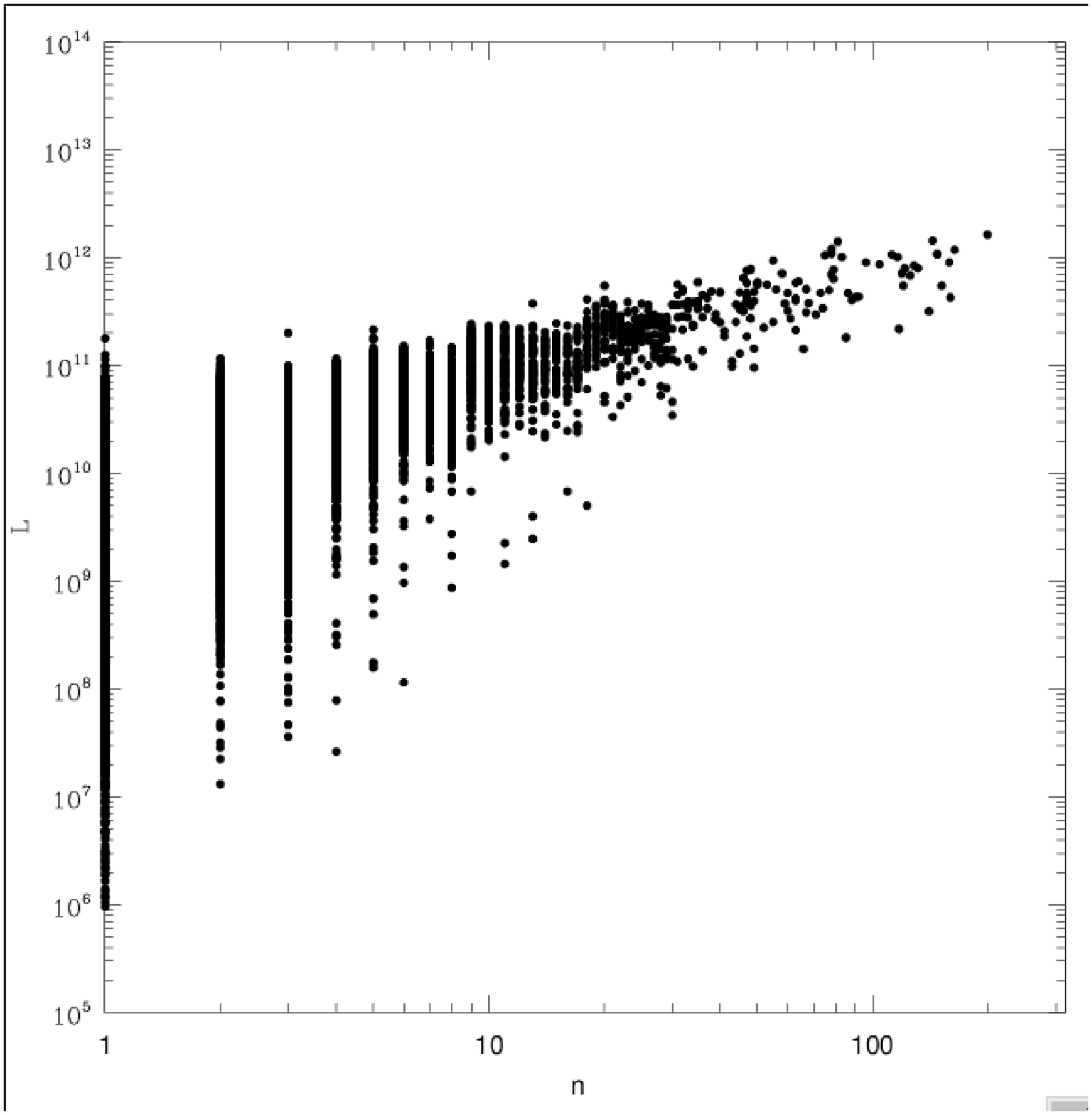}}
\caption{Luminosities, from the data catalogues, for all the groups as well as for the galaxies that do not belong to groups (n=1).} 
\label{fig:lum_n_data}
\end{center}
\end{minipage}
\hfill
\begin{minipage}{.45\textwidth}
\begin{center}
\centerline{\epsfxsize = 9.0cm
\epsfbox{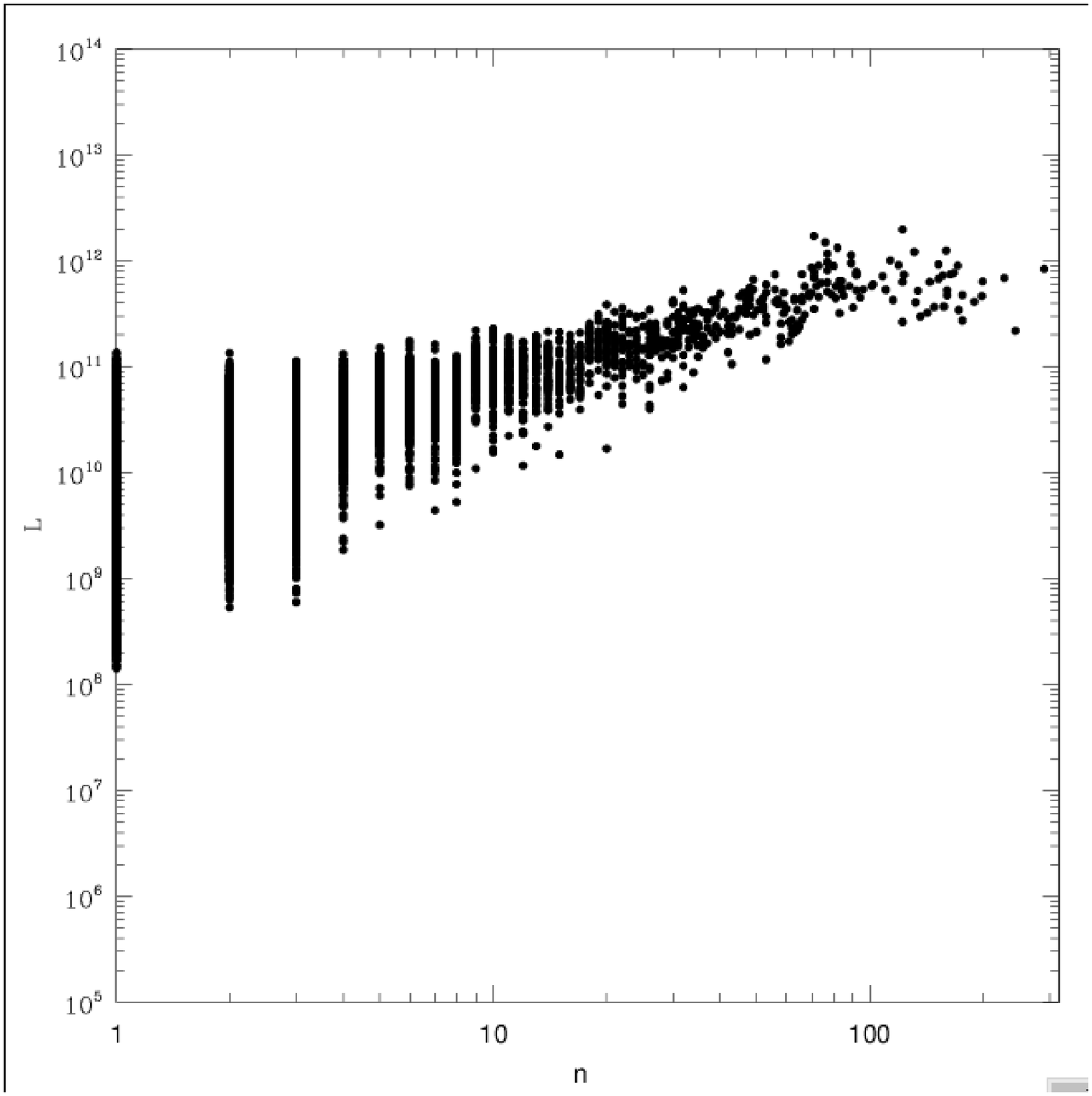}}
\caption{Luminosities, from the mock catalogues, for all the groups as well as for the galaxies that do not belong to groups (n=1)} 
\label{fig:lum_n_mock}
\end{center}
\end{minipage}
  \hfill
\end{figure*}

\begin{figure*}
\hfill
  \begin{minipage}{.45\textwidth}
\begin{center}
\centerline{\epsfxsize = 9.0cm
\epsfbox{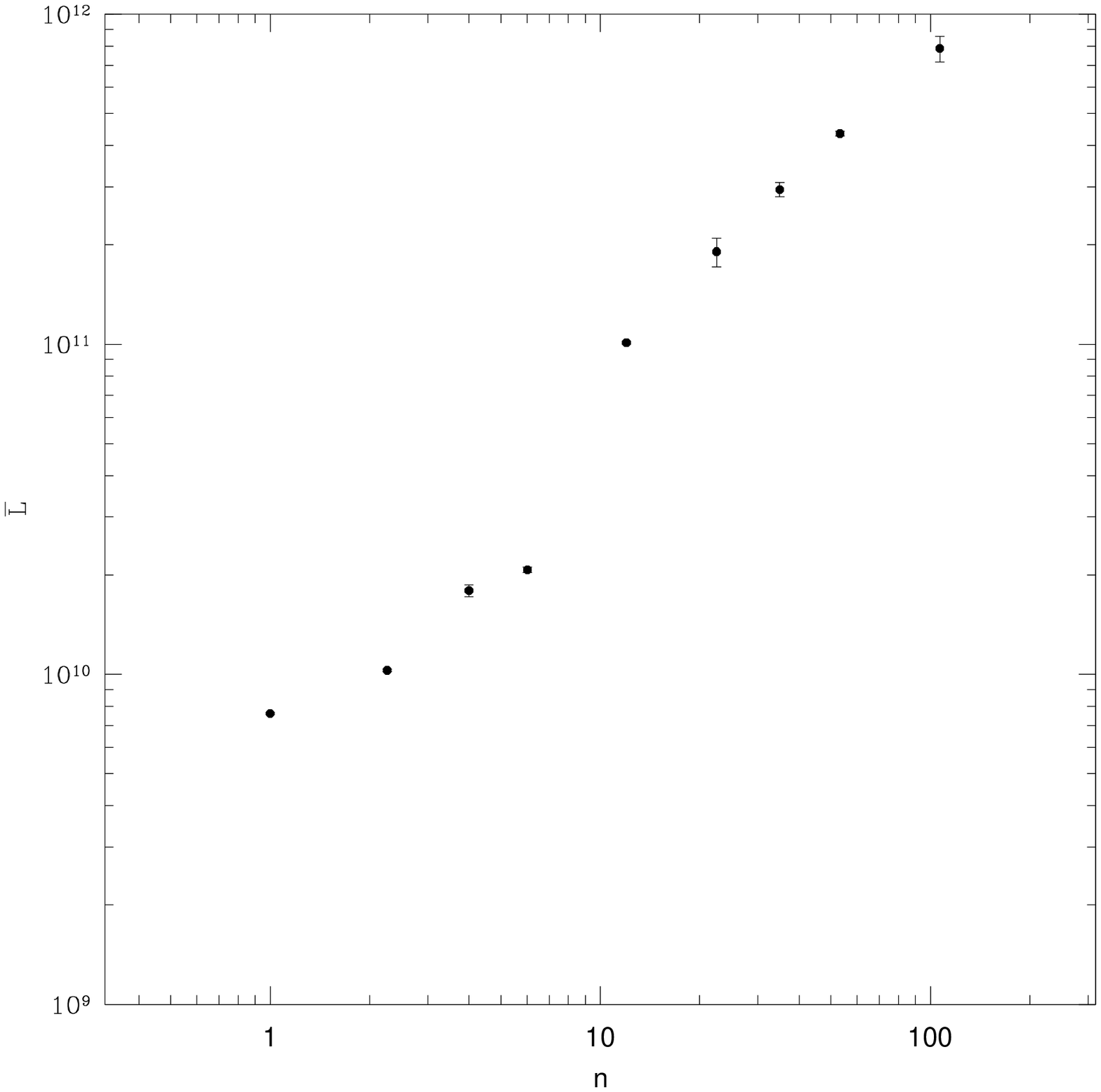}}
\caption{Average luminosities of our group samples (data).} 
\label{fig:ave_lum_n_data}
\end{center}
\end{minipage}
\hfill
\begin{minipage}{.45\textwidth}
\begin{center}
\centerline{\epsfxsize = 9.0cm
\epsfbox{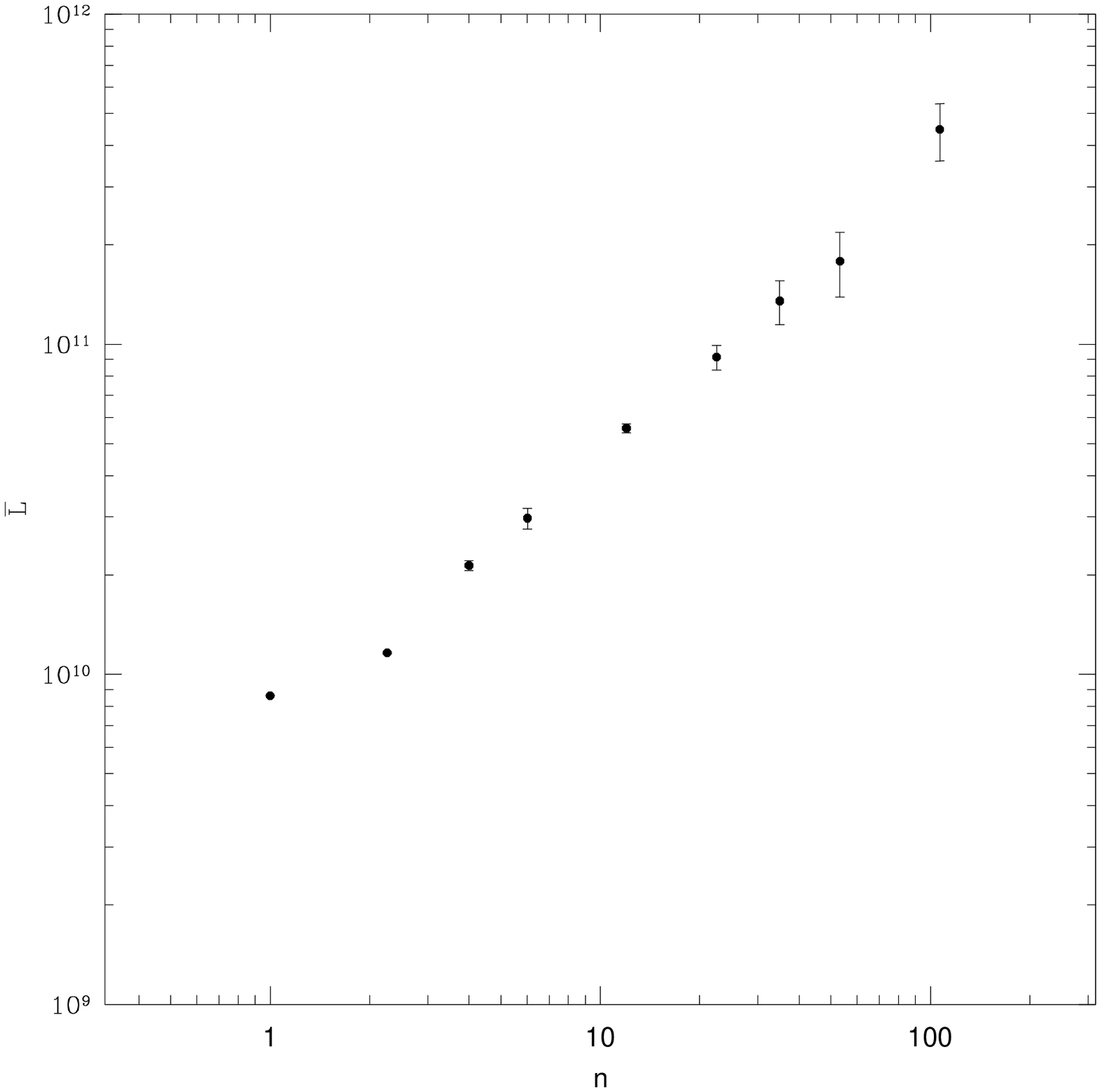}}
\caption{Average luminosities of our group samples (mock).} 
\label{fig:ave_lum_n_mock}
\end{center}
\end{minipage}
  \hfill
\end{figure*}

\subsection{Replacing the group membership with group luminosity} 

Figures \ref{fig:lum_n_data} and \ref{fig:lum_n_mock}  show the derived luminosities for all the groups for the data and the mocks respectively. Although, the ungrouped galaxies span a wide range of luminosities, especially in the case of the data, we notice that, as expected, bigger groups and clusters have bigger luminosities and their scatter is small. This can be better seen in Figures \ref{fig:ave_lum_n_data} and \ref{fig:ave_lum_n_mock} where we have calculated the average luminosity per group membership.

\begin{figure*}
\hfill
  \begin{minipage}{.45\textwidth}
\begin{center}
\centerline{\epsfxsize = 9.0cm
\epsfbox{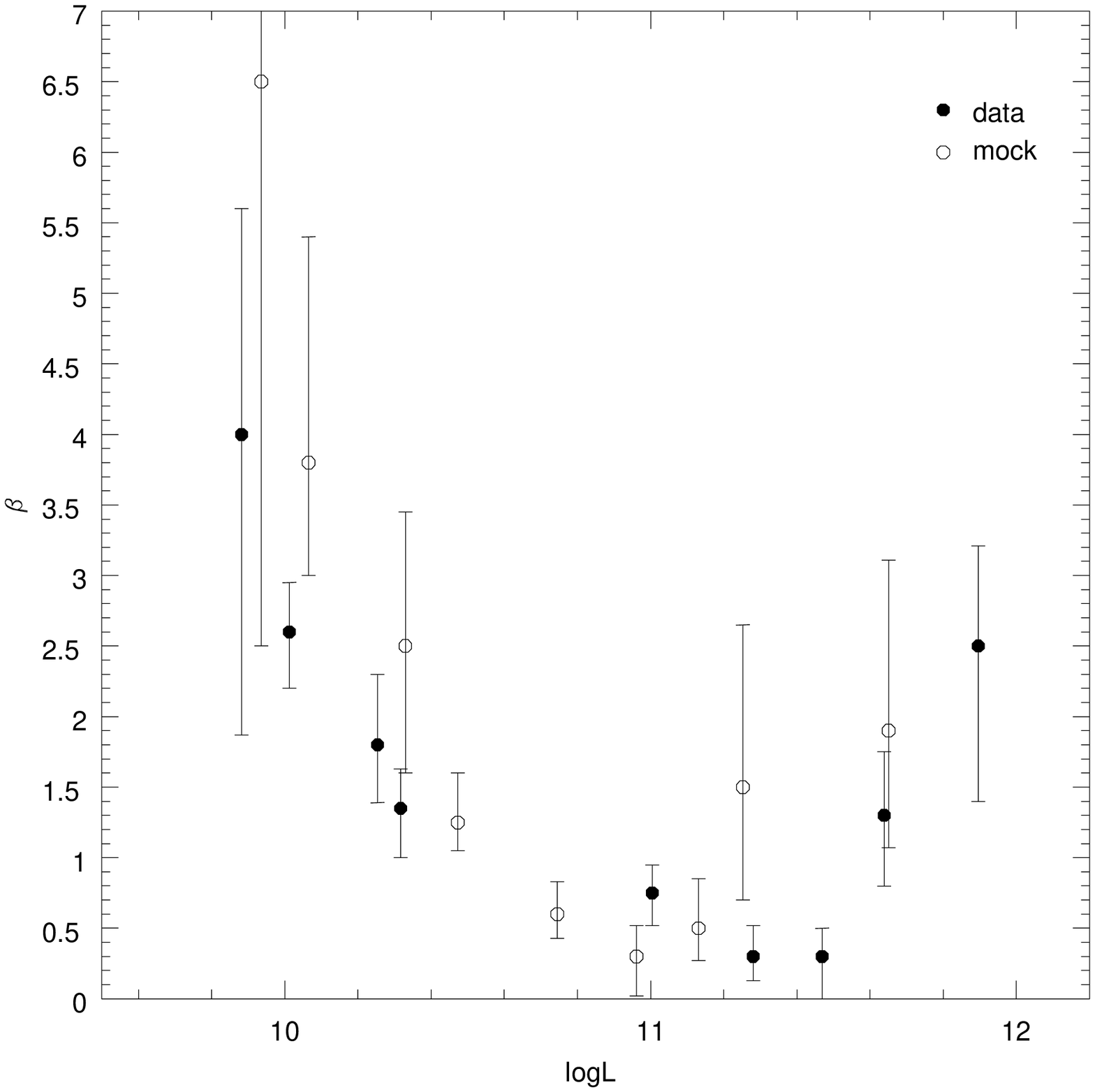}}
\caption{$\beta$ values for different group luminosities, when we substitute the average memberships with the corresponding luminosities.} 
\label{fig:b_vs_L_data_mock}
\end{center}
\end{minipage}
\hfill
\begin{minipage}{.45\textwidth}
\begin{center}
\centerline{\epsfxsize = 9.0cm
\epsfbox{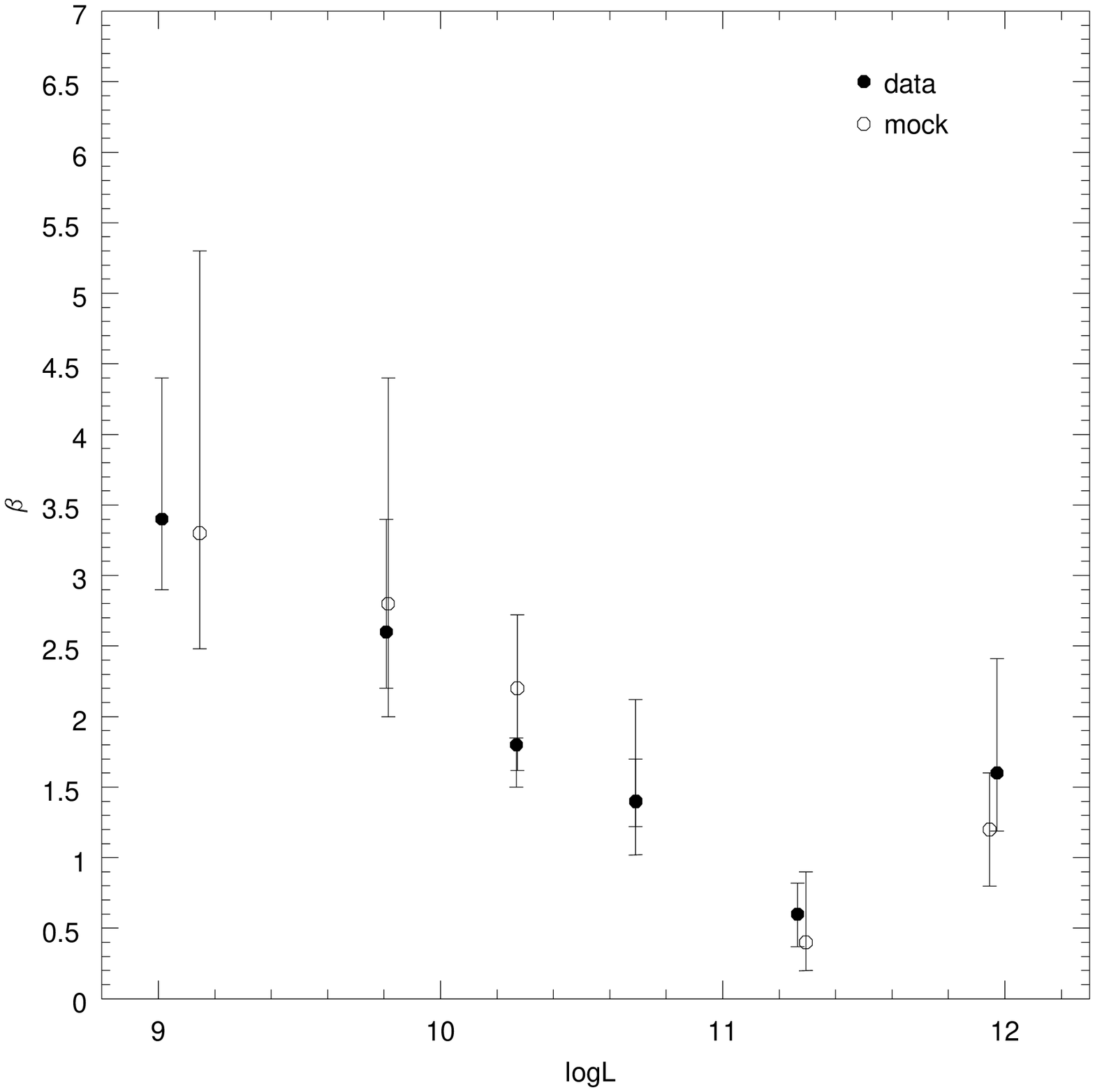}}
\caption{$\beta$ values for different group luminosity samples, when we re-sample our groups according to their luminosities.} 
\label{fig:b_vs_L_new}
\end{center}
\end{minipage}
  \hfill
\end{figure*}

Using now the average luminosity for each group sample, as shown in Figures \ref{fig:ave_lum_n_data} and \ref{fig:ave_lum_n_mock}, we replace the membership, n, with its corresponding observed luminosity. The results are shown in Fig \ref{fig:b_vs_L_data_mock} for the data and the mocks. The $\beta$ values shown in these Figures are those taken from the $\chi ^2$ velocity dispersions method described in the previous Section. The position of the minimum, which is what interests us, appears to be at lower luminosity for the mocks than for the data, although it may be claimed that the minimum is just flatter for the data. This difference is due to the fact that the mock catalogues contain less luminous intermediate groups and rich clusters than the data catalogues, as can be seen in Figures \ref{fig:ave_lum_n_data} and \ref{fig:ave_lum_n_mock}. Nevertheless, we can say that the minimum appears at $\approx10^{11}$h$^{-2}$L$_{\sun}$, i.e. about an order of magnitude higher than in Eke et al. results from $M/L$.

There are various reasons that may cause the difference between our results and those of Eke et al. Thus, in the next Section, in an attempt to better match the Eke et al. analysis, we shall re-sample our groups as a function of luminosity and estimate their M/L from velocity dispersions, in order to see if we can reproduce their results.

\begin{table}
\caption{Number of groups in the data and mock catalogues, when sampling them as a function of luminosity.}
\centering
\setlength{\tabcolsep}{4.0mm}
\begin{tabular}{lcc}
       \hline
group luminosity (h$^{-2}$L$_{\sun}$) & data & mock \\
       \hline \hline
$L\leq 2\times 10^9$  & $439$ & $99$ \\
       \hline
$2\times10^9<L\leq 10^{10}$ & $2,939$  & $2,311$ \\
       \hline
$10^{10}<L\leq 3\times 10^{10}$ & $7,421$  & $6,313$  \\
       \hline
$3\times10^{10}<L\leq 10^{11}$  & $4,332$  & $3,991$ \\
       \hline
$10^{11}<L\leq 6\times 10^{11}$ & $745$  & $681$ \\
       \hline
$L>6\times 10^{11}$ & $28$  & $45$  \\
       \hline 
\label{table:numbers_luminos}
\end{tabular}
\end{table}

\begin{table}
\caption{$s_0$ and $\gamma$ values for the redshift-space cross-correlation functions, $\xi _{cg} (s)$, using the luminosity based analysis.}
\centering
\setlength{\tabcolsep}{0.8mm}
\begin{tabular}{lcccc}
       \hline
\multicolumn{5}{c}{$\xi _{cg}(s)$ fits}\\
       \hline
$$& \multicolumn{2}{c}{data}& \multicolumn{2}{c}{mock} \\
       \hline
group luminosity (h$^{-2}$L$_{\sun}$) & $s_0$ & $\gamma$ & $s_0$ & $\gamma$ \\
       \hline \hline
$L\leq 2\times 10^9$ & $2.9\pm0.2$ & $1.8\pm0.1$ & $3.3\pm0.3$ & $1.6\pm0.1$\\
       \hline
$2\times10^9<L\leq 10^{10}$ & $3.3\pm0.1$ & $1.9\pm0.1$ & $3.9\pm0.1$ & $1.7\pm0.1$ \\
       \hline
$10^{10}<L\leq 3\times 10^{10}$  & $4.3\pm0.1$ & $1.5\pm0.1$ & $4.2\pm0.1$ & $1.5\pm0.1$ \\
       \hline
$3\times10^{10}<L\leq 10^{11}$ & $3.5\pm0.2$  & $1.7\pm0.1$ & $4.9\pm0.2$ & $1.5\pm0.1$  \\
       \hline
$10^{11}<L\leq 6\times 10^{11}$  & $8.5\pm0.2$  & $1.5\pm0.1$ & $7.1\pm0.2$ & $1.7\pm0.1$ \\
       \hline
$L>6\times 10^{11}$ & $14.8\pm0.8$  & $1.5\pm0.3$ & $10.2\pm0.4$ & $1.8\pm0.3$  \\
       \hline 
\label{fig:lum_xis_fits}
\end{tabular}
\end{table}

\begin{table}
\caption{Values of $<w_z^2>^{1/2}$ when we divide our group samples according to their luminosity ($\chi ^2$ method).}
\centering
\setlength{\tabcolsep}{1.8mm}
\begin{tabular}{lcccc}
       \hline
$$ & \multicolumn{2}{c}{data} & \multicolumn{2}{c}{mock} \\
group luminosity (h$^{-2}$L$_{\sun}$) & $fixed $ & $\chi ^2$ & $fixed $ & $\chi ^2$ \\
       \hline \hline
$L\leq 2\times 10^9$ & $102$ & $140$ & $130$ & $235$ \\
       \hline
$2\times10^9<L\leq 10^{10}$ & $135$ & $120$ & $144$ & $210$\\
       \hline
$10^{10}<L\leq 3\times 10^{10}$ & $166$ & $200$ & $169$ & $235$ \\
       \hline
$3\times10^{10}<L\leq 10^{11}$ & $206$ & $210$ & $209$ & $235$\\
       \hline
$10^{11}<L\leq 6\times 10^{11}$  & $345$ & $235$ & $383$ & $285$ \\
       \hline
$L>6\times 10^{11}$ & $601$ & $510$ & $652$ & $285$ \\
       \hline 
\label{table:w_lum}
\end{tabular}
\end{table}

\begin{table}
\caption{Values of $\beta$ when we divide our group samples according to their luminosity ($\chi ^2$ method).}
\centering
\setlength{\tabcolsep}{2.0mm}
\begin{tabular}{lcc}
       \hline
$$ & data & mock \\
       \hline 
group luminosity (h$^{-2}$L$_{\sun}$) & $\beta (\chi ^2) $ & $\beta (\chi ^2)$ \\
       \hline \hline
$L\leq 2\times 10^9$ & $3.40_{-0.50}^{+1.00}$  & $3.30_{-0.82}^{2.00}$  \\
       \hline
$2\times10^9<L\leq 10^{10}$ & $2.60_{-0.40}^{+0.80}$  & $2.80_{-0.80}^{+1.60}$ \\
       \hline
$10^{10}<L\leq 3\times 10^{10}$ & $1.80_{-0.30}^{+0.05}$  & $2.20_{-0.58}^{+0.52}$ \\
       \hline
$3\times10^{10}<L\leq 10^{11}$ & $1.40_{-0.38}^{+0.72}$  & $1.40_{-0.18}^{+0.30}$ \\
       \hline
$10^{11}<L\leq 6\times 10^{11}$  & $0.60_{-0.13}^{+0.02}$  & $0.40_{-0.02}^{+0.02}$ \\
       \hline
$L>6\times 10^{11}$ & $1.60_{-0.41}^{+0.81}$  & $1.20_{-0.12}^{+0.20}$  \\
       \hline 
\label{table:beta_lum}
\end{tabular}
\end{table}

\subsection{Sampling groups as a function of their luminosity}

According to Eke et al., group luminosity is the best way to rank groups in order of size especially for small groups where their luminosity can be better determined than their mass (Eke et al. 2004b, Figs 3 and 4). Therefore, we rank our group samples, not by membership as we did in Section 2, but by luminosity. We use 6 luminosity bins as shown in Table \ref{table:numbers_luminos}. Then for these new group samples we measure the redshift-space cross-correlation function using the methods described in Section 3 ($s_0$ and $\gamma$ values from the fits are shown in Table \ref{fig:lum_xis_fits}), the $\xi _{cg}(\sigma, \pi)$ cross-correlation function described in Section 6 and follow the fitting procedure described in Section 7 ($\chi ^2$ method), both for the data and the mock samples. We then obtain the values for the group-galaxy velocity dispersion, $<w_z^2>^{1/2}$, and infall parameter, $\beta$, shown in Tables \ref{table:w_lum} and \ref{table:beta_lum}, respectively. 

Fig. \ref{fig:b_vs_L_new} shows our new $\beta$ measurements. The agreement between the data and the mocks is still excellent and, as expected, we still find a minimum value for $\beta$. Nevertheless, the position of this minimum ($\approx10^{11}$h$^{-2}$L$_{\sun}$) continues to be significantly higher than that found by Eke et al. In order to check the consistency of our results, in the next Section we shall calculate the M/L ratio of our group samples.

\subsection{Mass-to-Light ratios}

We now want to see how the halo mass-to-light ratios vary as a function of group luminosity. Thus, we calculate the masses of the groups, using the expression (Eke et al. 2004a)

\begin{equation}
M=A\frac{\sigma ^2r}{G}
\end{equation}
where $\sigma$ is the group velocity dispersion (removing 85kms$^{-1}$ in quadrature for redshift errors, see previous Section), r is the rms projected galaxy separation of each group (Mpc/h), $A=5.0$, as used by Eke et at. 2004a and G is the gravitation constant. 

\begin{figure*}
\begin{center}
\centerline{\epsfxsize = 9.0cm
\epsfbox{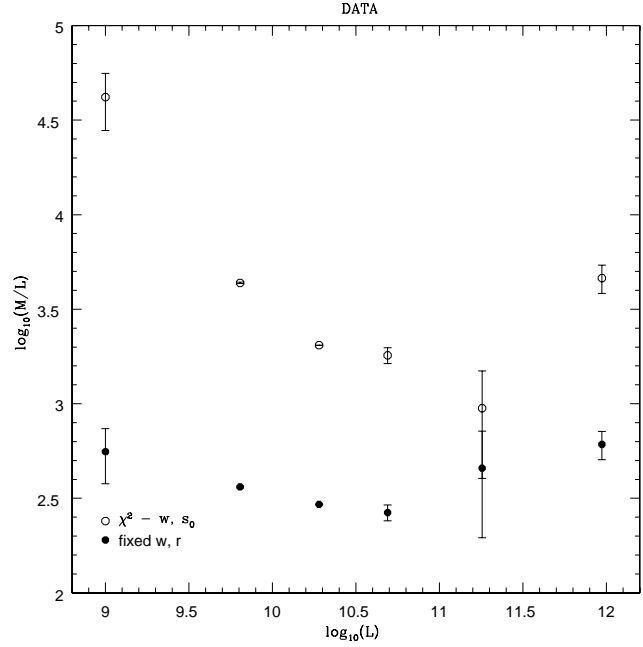}}
\caption{Average M/L ratio for each one of our data group luminosity samples. Filled circles show the results when we use the fixed values for the velocity dispersion and the values for the rms projected galaxy separation, as given by the 2PIGG team. Open circles show the average M/L ratio using our $\chi ^2$ measurements for the velocity dispersion of each group and the $s_0$ values estimated from fits to the redshift-space cross-correlation function.}
\label{fig:average_m_l_data}
\end{center}
\end{figure*}

\begin{figure*}
\begin{center}
\centerline{\epsfxsize = 9.0cm
\epsfbox{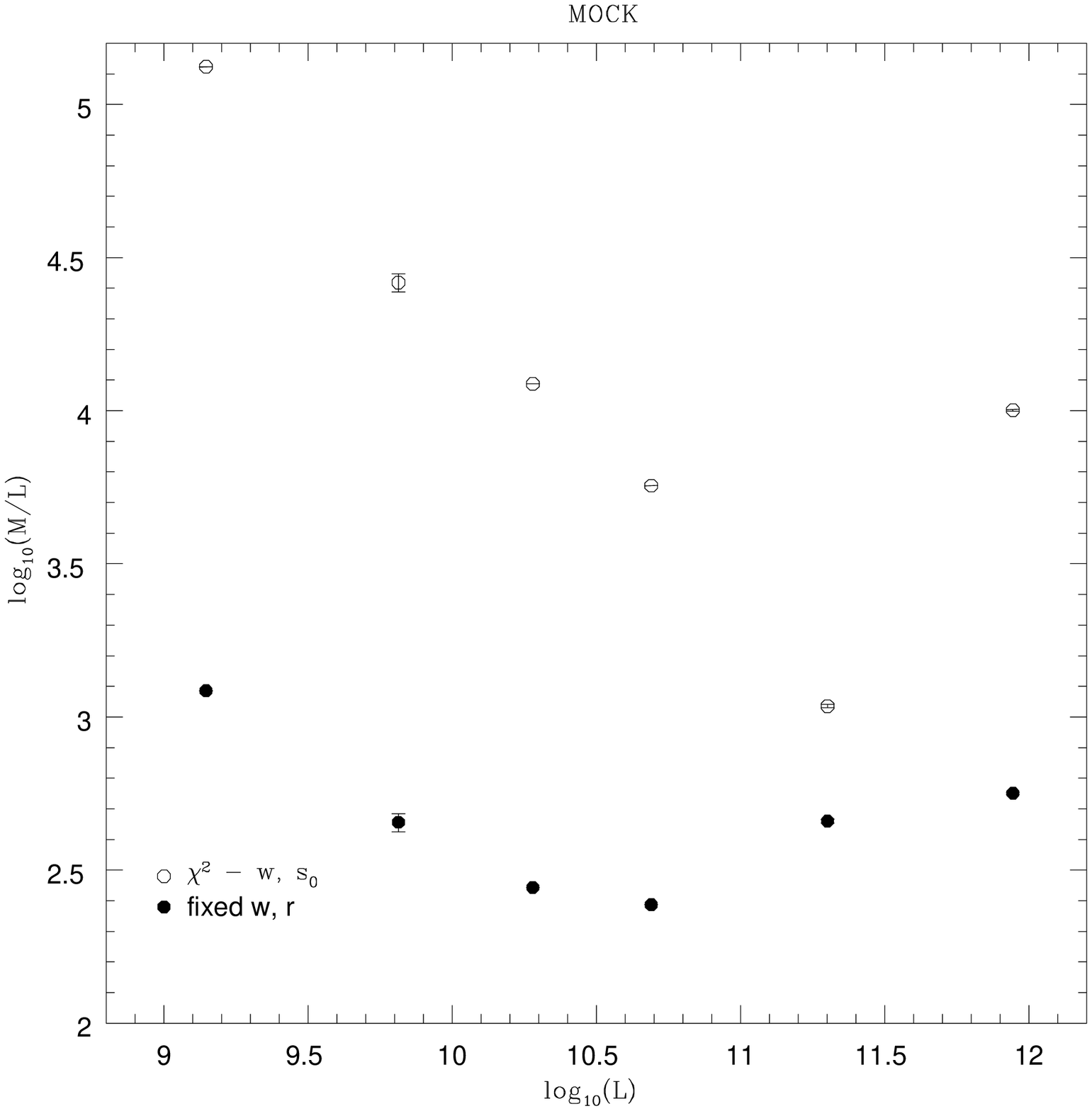}}
\caption{Average M/L ratio for each one of our mock group luminosity samples. Filled circles show the results when we use the fixed values for the velocity dispersion and the values for the rms projected galaxy separation, as given by the 2PIGG team. Open circles show the average M/L ratio using our $\chi ^2$ measurements for the velocity dispersion of each group and the $s_0$ values estimated from fits to the redshift-space cross-correlation function.}
\label{fig:average_m_l_mock}
\end{center}
\end{figure*}

We first use for $\sigma$, the average fixed velocity dispersions for each group luminosity sample from Eke et al. (2004a), as shown in Table \ref{table:w_lum} and the average r for each group sample from the rms projected galaxy separation values (Mpc/h) given by the 2PIGG team. Then we repeat the estimates, replacing r by $s_0$ (Table \ref{fig:lum_xis_fits}) and the fixed velocity dispersion by the $\chi ^2$ fit, for $\sigma$ (Table \ref{table:w_lum}). The two estimates are shown in Figures \ref{fig:average_m_l_data} and \ref{fig:average_m_l_mock} for the data and the mock catalogues, respectively. As we see, data and mocks are in excellent agreement. The difference in the M/L ratios (open and filled circles) is because the $s_0$ values are higher than the r values. The most important feature of these Figures is the behaviour of the position of the minimum. Using the results from our analysis ($s_0$ and $<w_z^2>^{1/2}$, open circles), in agreement with our previous results for $\beta$, the minimum appears at $L\simeq2\times10^{11}$h$^{-2}$L$_{\sun}$ whereas using the fixed $<w_z^2>^{1/2}$ and the r values from the 2PIGG team (filled circles) it appears at lower luminosities, i.e. $L\simeq5\times10^{10}$h$^{-2}$L$_{\sun}$. This difference at the position of the minimum of the M/L ratio, is because the fixed values for $<w_z^2>^{1/2}$ show a "jump" for the two most luminous group samples, resulting in larger masses for those whereas the values for $<w_z^2>^{1/2}$ using the $\chi ^2$ minimisation remains roughly the same for all the group samples, as shown in Table \ref{table:w_lum}. Therefore, the minimum using these fixed $<w_z^2>^{1/2}$ values appears at our third most luminous group rather the second.

\begin{figure*}
\begin{center}
\centerline{\epsfxsize = 9.0cm
\epsfbox{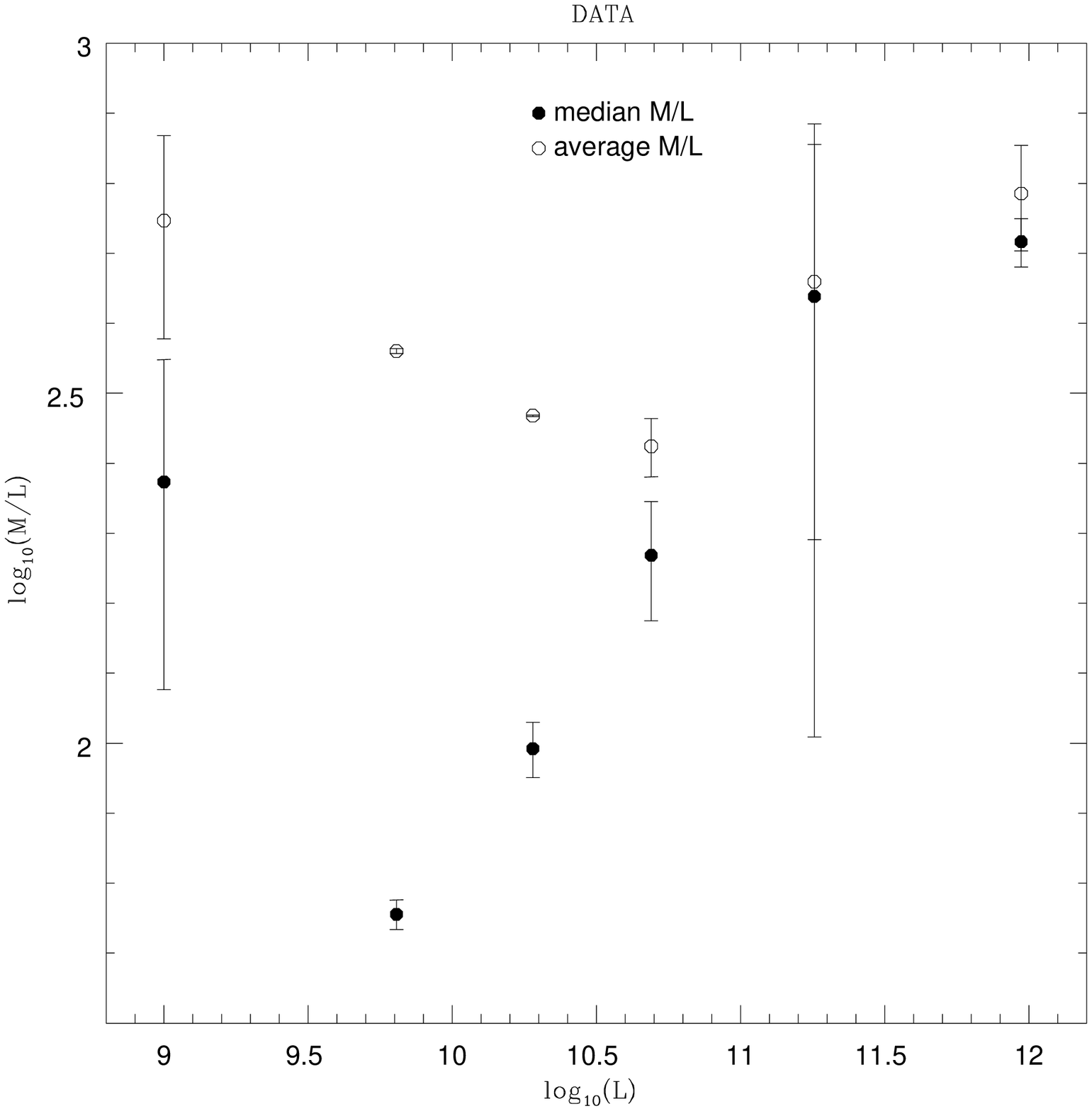}}
\caption{A comparison between the average (open circles) and the median (filled circles) values of the M/L ratio for each one of our data group luminosity samples. The median values move the minimum of the ratio to $10\times$ lower group luminosities.}
\label{fig:median_m_l_data}
\end{center}
\end{figure*}

\begin{figure*}
\begin{center}
\centerline{\epsfxsize = 9.0cm
\epsfbox{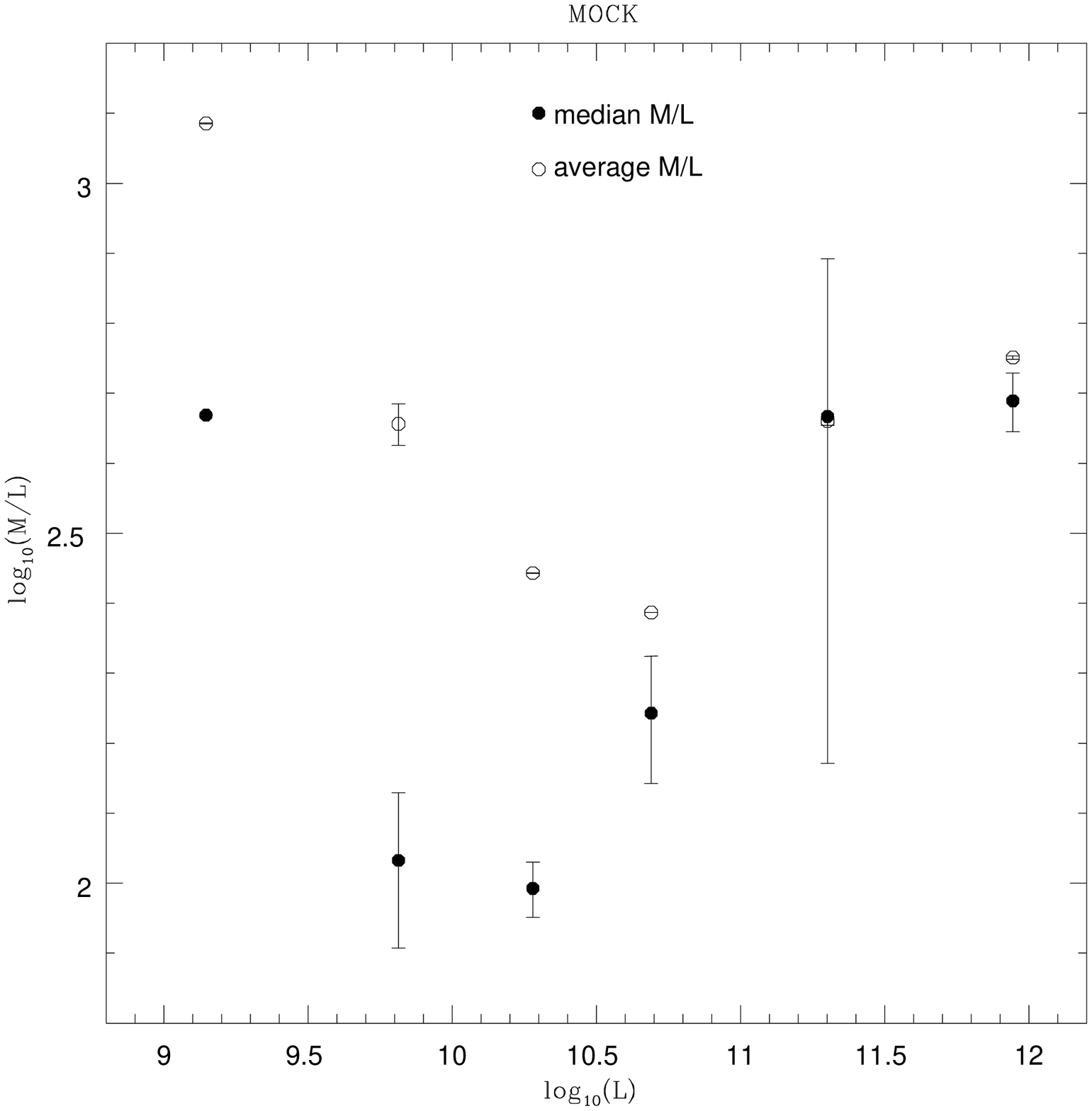}}
\caption{A comparison between the average (open circles) and the median (filled circles) values of the M/L ratio for each one of our mock group luminosity samples. The median values move the minimum of the ratio to $3\times$ lower group luminosities.}
\label{fig:median_m_l_mock}
\end{center}
\end{figure*}

Although using the Eke et al. velocity dispersion and group sizes has reduced the discrepancy in the position of the $M/L$ minimum by a factor of $\sim 3$ a difference of a factor of 8 between the $M/L$ and $\beta$ minima still persists. Investigating this further we now use, instead of average mass and luminosity values for each group, their median values. The results are shown in Figures \ref{fig:median_m_l_data} and  \ref{fig:median_m_l_mock} for the data and the mock, respectively. We notice that there are differences between the two methods. The average values have slightly higher M/L values but most importantly the minimum in the M/L ratio, when we use the median values, appears at smaller luminosities. Actually now, comparing those two Figures with Fig. 15 in Eke et al. 2006 for the data, and Fig. 16 in Eke et al. 2004b for the mocks we see that we reproduce all the features of the plots of the 2PIGG team. More specifically, M/L ratio increases by a factor of 5 when spanning luminosities from $10^{10}-10^{12}$h$^{-2}$L$_{\sun}$ and shows a minimum at $\approx5\times10^9$h$^{-2}$L$_{\sun}$ in excellent agreement with the position of the minimum as found by the 2PIGG team.

\subsection{Reasons for the difference in $\frac{M}{L}$ and $\beta$ minima}

Summarising the reasons for the difference in the position of the minima between $\beta (L)$ and $\frac{M}{L}(L)$, we note that there are already different minima shown by $\frac{M}{L}(L)$ depending on whether median or average $\frac{M}{L}(L)$ values are used. When the median $\frac{M}{L}(L)$ is used, the position of the minimum decreases by a factor of $\sim 8$ in L. These conclusions also apply when assuming the velocity dispersions of Eke et al. instead of our $\chi ^2$ fitted values ($3\times$ lower). Thus, if the $\chi ^2$ fitted velocity dispersion and average $M/L$ are used the position of the minimum of the $M/L$ moves up in L and appears at the same luminosities as in $\beta$. Eke et al. claim that the simulations suggest the use of the median in the velocity dispersion fits. If the simulations are correct then this would imply that the position of the $\beta$ minimum in L is either due to other physical effects or that we shouldn't be using average values for $\beta$. The other physical effects include the suggestion that the $\frac{M}{L}$ and $\beta$ results may apply to different scales around the cluster. There also remain questions as to whether it is fair to compare $\frac{M}{L}$ estimates with with $\beta\sim\frac{1}{b}\sim\delta \rho _{mass}/\delta \rho _{galaxies}$.

\section{Discussion $+$ Conclusions}

Using the 2PIGG galaxy group catalogue we have investigated the behaviour of the infall parameter $\beta$ for group-galaxies of different membership, as well as their rms velocity dispersions via the redshift distortion of the group-galaxy cross-correlation functions. 

We first separated our galaxy group set into subsamples, following approximately the Eke et al. 2004a membership classes. We also used mock catalogues, created by the 2PIGG team, and we applied the same membership classification to them, in order to compare our results from the data sets with those from the mock catalogues.  

We cross-correlated these group samples with 2dFGRS galaxies, in order to measure the redshift-space, $\xi _{cg} (s)$, the semi-projected, $w_p(\sigma)/\sigma$ and the real-space, $\xi _{cg} (r)$, cross-correlation functions. For each of them, we fitted a power law, e.g. $\xi _{cg}(r)=(r/r_0)^{-\gamma}$. In agreement with previous studies, we noticed that the correlation amplitude increased, both for the data and the mocks, as we moved to richer groups. Also, the redshift-space cross-correlation amplitude and slope are statistically in agreement with those from the real-space cross-correlation function. Comparing the data and the mock for each of the cross-correlation functions, we noticed that, they are all in agreement, especially the results from the redshift-space function, which is the least noisy.

Next, we measured the $\xi _{cg} (\sigma, \pi)$ cross-correlation function and used its shape and the $s_0$ and $\gamma$ from the $\xi _{cg} (s)$ fits to model the redshift-space distortions and measured the $\chi ^2$ group-galaxy rms velocity dispersion, $<w_z^2>^{1/2}$, and the infall parameter, $\beta$. Our measurements showed that $<w_z^2>^{1/2}$ remains roughly constant for small and intermediate group-galaxies and rises for clusters whereas the values of $<w_z^2>^{1/2}$, estimated by the 2PIGG team increase with increased group membership. The $\beta$ results for the data and the mocks are in very good agreement and show a minimum for the dynamical infall at intermediate group memberships. 

Prompted by previous studies of group M/L ratios as a function of their luminosities, in the last Section we calculated the average luminosities for each of our group samples and replaced the group memberships with their corresponding luminosities. This revealed a discrepancy of more than an order of magnitude in the position of the minimum between $\beta$ and $M/L$. In order to examine this discrepancy we re-sampled our groups using as a criterion their luminosity instead of their membership and calculated their M/L ratios. This analysis revealed that the reasons for the difference in the position of this minimum is due to the different velocity dispersion measurements between us and Eke et al. and most importantly due to the use of median instead of average values. 

Our overall conclusion is that bias estimates from dynamical infall broadly appear to support the minimum in star-formation efficiency at intermediate halo masses and also that there may be slight differences with mock catalogues. However, there is a systematic shift between the $M/L$ and $\beta (\propto \frac{1}{b}$) minima. Judged by the mock results, the use of the median values than the average seems to give more accurate $M/L$ results. Unfortunately, there is no option to use a median rather than an average $\beta$ for the z-distortion results. There is also the possibility that the mock $M/L$ with L has been misestimated by Eke et al. 2004b, since this involves friends-of-friends routines applied to define halo masses and then independently to find group luminosities and the process of matching the two is not without its complications. More work is therefore needed on the DM mocks to check further the reasons for this difference in interpretation. However, it may be more likely that our larger scale estimates of $\beta$ measure a different aspect of the galaxy mass-luminosity relation than the small scale $\sigma$ measurements. Also there may remain issues about the validity of comparing the dynamical infall parameter $\beta (\sim\frac{1}{b}\sim\delta \rho _{mass}/\delta \rho _{galaxies})$ with the $\frac{M}{L}$ ratio from the velocity dispersions.

Finally, the results presented here seem to be in agreement with our previous QSO-galaxy group lensing results (Myers et al. 2003, Mountrichas \& Shanks 2007). There we found that $n\geq 7$ groups showed large effective masses and galaxies also showed an effective anti-bias of $b\approx 0.2$ or $\beta \approx 2.5$. These $n\geq 7$ groups have luminosities which places them at $\sim5\times10^{10}L\sun$. In Fig. \ref{fig:b_vs_L_new} they, thus, would appear to have a higher $\beta$ than those at minimum, in agreement with the lensing conclusions. Moreover individual galaxies appear to have $\beta \sim3$ or $b\sim0.15$ again in agreement with the conclusion for the galaxy lensing results of Myers et al. 2005. The only surprise is that the mock catalogues show similar behaviour. This suggests that such large $\beta$ variations might be expected even in standard galaxy formation models. 

\section{Acknowledgments}

The 2dFGRS has been made possible through the dedicated efforts of the staff of the Anglo-Australian Observatory, both in creating the 2dF instrument and in supporting it on the telescope. We would also like to thank the 2PIGG team for making available their group data and mock catalogues.

\vspace{10 mm}

\noindent
{\bf References}

\vspace{6 mm}

\noindent
Ballinger, W. E., Peacock, J. A., Heavens, A. F., 1996, MNRAS, 282, 877

\vspace{3 mm}

\noindent
Beers, T. C., Flynn, K., Gebhardt, K., 1990, AJ, 100, 32B

\vspace{3 mm}

\noindent
Borgani, S., Plionis, M., Kolokotronis, V., 1999, MNRAS, 305, 866

\vspace{3 mm}

\noindent
Cole, S., Lacey, C. G., Baugh, C. M., Frenk, C. S., 2000, MNRAS, 319, 168

\vspace{3 mm}

\noindent
Croft, R. A. C., Dalton, G. B., Efstathiou, G., Sutherland, W. J., Maddox, S. J., 1997, MNRAS, 291, 305

\vspace{3 mm}

\noindent
da $\hat{A}$ngela, J. et al., 2005, astro-ph/0612401, submitted to MNRAS

\vspace{3 mm}

\noindent
da $\hat{A}$ngela, J., Outram, P. J., Shanks, T., 2005, MNRAS, 361, 879

\vspace{3 mm}

\noindent
da $\hat{A}$ngela, J., Outram, P. J., Shanks, T., Boyle, B. J., Croom, S. M., Loaring, N. S.; Miller, L., Smith, R. J. 2005, MNRAS, 360, 1040

\vspace{3 mm}

\noindent
Efstathiou, G., 1988, Proceedings of the Third IRAS Conference, Queen Mary College, London, England, July 6-10, 1987. Lecture Notes in Physics, Vol. 297, edited by A. Lawrence. Springer-Verlag, Berlin, 1988., p.312

\vspace{3 mm}

\noindent
Eke V. R., et al., 2004, MNRAS, 348, 866

\vspace{3 mm}

\noindent
Eke V. R., et al., 2004, MNRAS, 355, 769

\vspace{3 mm}

\noindent
Eke V. R., Baugh C. M., Cole S., Frenk C. S., Navarro J. F., 2006, MNRAS, 370, 1147

\vspace{3 mm}

\noindent
Giuricin, G., Samurovi\u0107, S., Girardi, M., Mezzetti, M., Marinoni, C., 2001, ApJ, 554, 857G

\vspace{3 mm}

\noindent 
Hawkins et al., 2003, MNRAS, 346, 78

\vspace{3 mm}

\noindent
Hoyle, F., Outram, P. J., Shanks, T., Boyle, B. J., Croom, S. M., Smith, R. J., 2002, MNRAS, 332, 311

\vspace{3 mm}

\noindent
Jenkins, A., Frenk, C. S., Pearce, F. R., Thomas, P. A., Colberg, J. M., White, S. D. M., Couchman, H. M. P., Peacock, J. A., Efstathiou, G., Nelson, A. H., 1998, ApJ, 499, 20

\vspace{3 mm}

\noindent
Kaiser N., 1987, MNRAS, 227, 1

\vspace{3 mm}

\noindent
Lewis, I. J., et al. 2002, MNRAS, 333, 279L

\vspace{3 mm}

\noindent
 Loveday, J., Maddox, S. J., Efstathiou, G., Peterson, B. A., 1995, ApJ, 442, 45L

\vspace{3 mm}

\noindent
Mountrichas, G., Shanks T., 2007, MNRAS, 380, 113M

\vspace{3 mm}

\noindent
Mountrichas, G., Shanks T., 2008 in prep.

\vspace{3 mm}

\noindent
Myers A. D., Outram P. J., Shanks T., Boyle B. J., Croom S. M., Loaring N. S., Miller L., Smith R. J., 2003, MNRAS, 342, 467

\vspace{3 mm}

\noindent
Myers A.D., Shanks T., Boyle ÊB.ÊJ., Croom ÊS.ÊM., Loaring ÊN.S., 
Miller ÊL. \& Smith ÊR.J. 2005, MNRAS, 359, 741.

\vspace{3 mm}

\noindent
Outram, P. J.; Hoyle, Fiona; Shanks, T., 2001, MNRAS, 321, 497

\vspace{3 mm}

\noindent
Padilla, N. D., Merchán, M. E., Valotto, C. A., Lambas, D. G., Maia, M. A. G., 2001, ApJ, 554, 873P

\vspace{3 mm}

\noindent
Padilla, N. D., Merchán, M. E., Valotto, C. A., Lambas, D. G., Maia, M. A. G., 2001, astro-ph/0102372

\vspace{3 mm}

\noindent
Ratcliffe, A., Shanks, T., Broadbent, A., Parker, Q. A., Watson, F. G., Oates, A. P., Fong, R., Collins, C. A., 1996, MNRAS, 281L, 47R

\vspace{3 mm}

\noindent
Ratcliffe, A., Shanks, T., Parker, Q. A., Fong, R., 1998, MNRAS, 296, 191

\vspace{3 mm}

\noindent
Robotham, A., Wallace, C., Phillipps, S., De Propris, R., 2006, ApJ, 652, 1077R

\vspace{3 mm}

\noindent
Ross, N. P., Shanks, T., Cannon, R. D., Wake, D. A., Sharp, R. G., Croom, S. M., Peacock, John A., astro-ph/0704.3739, submitted to MNRAS

\vspace{3 mm}

\noindent
Saunders, W., Rowan-Robinson, M., Lawrence, A., 1992, MNRAS, 258, 134

\vspace{3 mm}

\noindent
Taylor, A. N., Ballinger, W. E., Heavens, A. F., Tadros, H., 2000, 2001, MNRAS, 327, 689T

\vspace{3 mm}

\noindent
Willmer, C. N. A., da Costa, L. N., Pellegrini, P. S., 1998, AJ, 115, 869

\vspace{3 mm}

\noindent
Yang, X., Mo, H.J., van den Bosch, F. C., Jing, Y. P., 2005a, MNRAS, 356, 1293

\vspace{3 mm}

\noindent
Yang, X., Mo, H.J., van den Bosch, F. C., Jing, Y. P., 2005b, MNRAS, 357, 608

\vspace{3 mm}

\noindent
Yang, X., Mo, H.J., van den Bosch, F. C., Jing, Y. P., 2005c, MNRAS, 362, 711

\vspace{3 mm}

\noindent
Yang, X., Mo, H.J., van den Bosch, F. C., Weinmann, S. M., Cheng, L., Jing, Y. P., 2005d, MNRAS, 362, 711

\vspace{3 mm}

\end{document}